\documentclass[journal]{IEEEtran}
\usepackage{comment}
\usepackage{amssymb, amsmath,graphicx,charter, latexsym}
\usepackage{amsfonts}
\usepackage{bbm}
\newtheorem{definition}{Definition}
\usepackage{enumerate}
\newtheorem{lemma}{Lemma}
\newtheorem{corollary}{Corollary}

\newtheorem{theorem}{Theorem}

\newtheorem{remark}{Remark}

\usepackage{graphicx}
\usepackage{url}
\usepackage{epstopdf}
\usepackage{algorithmic}
\usepackage{algorithm}
\usepackage{xcolor}
\usepackage[fleqn]{mathtools}
\usepackage[noabbrev]{cleveref}
\begin{document}
\title{Optimal Decentralized Dynamic Policies for Video Streaming over Wireless Channels}

\author{Rahul Singh~\IEEEmembership{Member,~IEEE,} and
        P.~R.~Kumar~\IEEEmembership{Fellow,~IEEE,}
        }

\maketitle
\IEEEpeerreviewmaketitle

\begin{abstract}
The problem addressed is that of optimally controlling, in a decentralized fashion, the download
of mobile video, which is expected to 
comprise 75\% of total mobile data traffic by 2020.
The server can dynamically choose which
packets to download to clients, from among several packets which encode their videos at different resolutions,
as well as the power levels of their transmissions. This allows it to
control packet delivery probabilities, and thereby, for example, avert imminent video outages at clients.
It must however respect the access point's constraints on bandwidth and
average transmission power. The goal is to maximize video ``Quality of Experience" (QoE),
which depends on several factors such as
(i) outage duration
when the video playback buffer
is empty, 
(ii) number of
outage periods, 
(iii) how many frames downloaded are of lower
resolution, 
(iv) temporal variations in resolution, etc.

It is shown that there exists an optimal decentralized solution where the AP announces the price of energy,
and each client distributedly and dynamically maximizes its own QoE subject to the cost of energy. A distributed iterative algorithm to solve for optimal decentralized policy is also presented.
Further, for the client-level QoE optimization, the optimal choice of video-resolution and power-level of packet transmissions
has a simple monotonicity and threshold structure vis-a-vis video playback buffer level. 
When the number of orthogonal channels is less than the number of clients, 
there is an index policy for prioritizing packet transmissions.
When the AP has to simply choose
which clients' packets to transmit, the index policy is asymptotically optimal
as the number of channels is scaled up with clients. 
\end{abstract}

\begin{IEEEkeywords}
Quality of Experience, Video Streaming, Video on Demand, Video Download, Wireless Networks.
\end{IEEEkeywords}
\section{Introduction}
Mobile video traffic accounted for 55$\%$ of total mobile data traffic in 2015, and its dominance
is expected to increase to 75$\%$ by 2020. 
Optimally supporting such video downloads from an access point (AP) requires:
\begin{enumerate}[i)]
\item adaptively choosing the bit-rates of the
variable bit-rate encoded videos for several clients, and
\item the transmission powers of packets to the clients,
\item
according to the time-varying wireless channels of the several clients, and
\item according to levels of the video playback buffers of the clients,
\item in such a way as to maximize ``Quality of Experience" (QoE)
that is a complex composite of simpler Quality of Service (QoS) metrics such as throughput, delay and outages,
as well as more complicated factors such as frequent switchings between differing resolutions,
\item while taking constraints such as total access point transmit power into account.
\end{enumerate}
Due to the random nature of wireless, it is a stochastic system. It is a decentralized control system since individual agents (video clients) can only observe their own variables and states (such as their own video buffer content) when making decisions (such as the resolution of the packet requested, and the power level at which it is transmitted). Control constraints arise since the video has power constraints. From a control theoretic point of view, this therefore gives rise to a decentralized stochastic control problem with multiple agents, with long-term average constraints on the controls of agents. For such problems, in general, there need not exist an optimal policy that is decentralized. Further, in general, there is also the computational challenge of determining the optimal policy with a tractable amount of computation.  \par 

\begin{figure}[!t]
	\centering
	\includegraphics[width=0.5\textwidth]{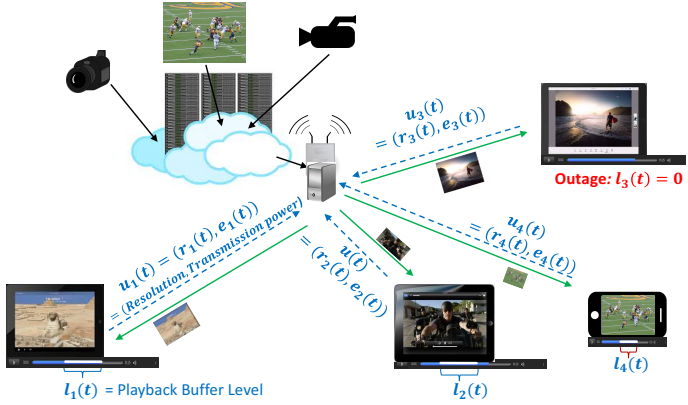}
	\caption{Under the proposed decentralized optimal solution, each client $n$ at time $t$ requests a packet containing the optimal resolution $r_n(t)$ of the video, at an optimal transmission power $e_n(t)$, which requires choosing an optimal decentralized feedback control $u_n(t)=(r_n(t),e_n(t))$ based on its instantaneous playback buffer level $\ell_n(t)$, wireless channel condition, and a price of energy announced by the access point.}
	\label{figsoln}
\end{figure}

This paper addresses how to perform video downloads optimally, and in a tractable manner.
It presents a decentralized solution that allows each client to make its own decisions
on choosing the resolutions and transmit power levels of packets based on its own video playback buffer level and a global price of
energy, as shown in Figure~\ref{figsoln}.
\begin{figure}[!t]
	\centering
	\includegraphics[width=0.5\textwidth]{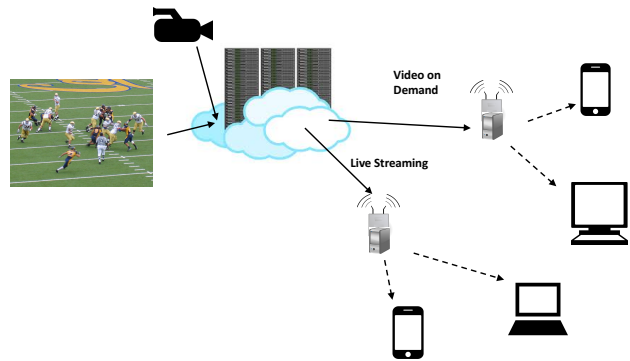}
	\caption{A Cloud live-streaming a game, and hosting a Video on Demand (VoD) service. A continuous adaptation of bit rate of \emph{each} user is required in order to ensure a high level of Quality of Experience to end users.}
	\label{figcloud}
\end{figure}

Traditional well-studied QoS metrics such as throughput and delay are of little use for judging user experience for video streaming applications. The QoE associated with video streaming depends on several complex metrics~\cite{Balachandran:2013,Dobrian:2011,Shafiq:2014}.
In order to meet stringent QoE requirements, service providers have switched to advanced platforms such as Cloud based services~\cite{5754008} and Content Delivery Networks (CDNs)~\cite{Pallis:2006:IPC:1107458.1107462}, which utilize adaptive bitrate streaming algorithms such as DASH~\cite{dash} and HTTP Live Streaming (HLS) in order to continually monitor and improve the streaming experience.
A party subscribing to a live video streaming service from popular cloud service such as Microsoft Azure, IBM Cloud, Google Cloud, Amazon CloudFront, Apple's iCloud generates a video file and uploads it to the cloud in real-time; see Fig.~\ref{figcloud}. The cloud then transcodes this data into multiple bit-rates, and the audience of this stream is served the video file using adaptive streaming algorithms such as DASH. DASH enables a viewer to switch to low resolution video in case its connection bandwidth is reduced, thus avoiding video interruptions. Since a major chunk of video data is demanded by mobile devices that typically have bandwidth fluctuations, this enables the streaming service to reach a wider range of audiences. 

However, state-of-the art adaptive streaming algorithms are unable to provide a satisfactory QoE for video streaming. As an example, the popular DASH algorithm is either too slow to respond to changes in congestion levels, or is overly sensitive to short-term network bandwidth variations~\cite{a5}. Similarly, for clients served over wireless networks~\cite{a8}, rate adaptation needs to take complex factors such as channel fading into account while making streaming decisions. Experimental studies of rate adaptation techniques employed by popular DASH clients such as Microsoft Smooth Streaming~\cite{microsoft}, Adobe OSMF~\cite{adobe}, and Netflix have demonstrated that these algorithms perform poorly.

The video streaming experience of a client is determined by several metrics such as a) outage probability, i.e., the average time spent without video streaming due to non-availability of video packets; b) time spent in rebuffering packets; c) average video quality~\cite{Dobrian:2011}; e) temporal variations in resolutions of video packets~\cite{yim}. Hence, a scheduling policy that is designed to maximize the QoE needs to simultaneously achieve optimal trade-offs between several of these metrics. For example, streaming low resolution packets reduces the outage probability since low-resolution packets are associated with lower bit-rates or equivalently higher probability of successful packet transmission over an unreliable channel. However, lower resolution packets also reduce the average QoE, and hence the AP would like to switch to streaming higher resolution packets opportunistically. But then, the action of switching between different resolutions also introduces temporal variations in video quality, which in turn reduces the QoE. In addition, if the channel gain of the wireless channels connecting the AP to clients is time-varying, then one also needs to take into account the dynamics of the wireless channel~\cite{Shafiq:2014} while making scheduling decisions. 
For example, the algorithm could switch to a low resolution video upon detecting a reduction in the bandwidth of a client. 
Yet another control variable is the transmission power of packets, which affects the success of the transmission.
In addition, the access point (AP) usually has some form of constraint on bandwidth, such as the average power that it can consume, or the number of clients that it can schedule simultaneously. 
The resulting overall dynamic optimization could be delegated to a centralized controller but it would need to have knowledge of the states of all clients,
such as their instantaneous playback buffer levels at each time $t$ or their channel states. Moreover, the computation of such
an optimal centralized control policy suffers from the curse of dimensionality since the state space scales as $B^{N}$, where $B$ is the bound on playback buffer size of clients and $N$ is the number of clients.

Our main results are as follows. Though, in general, for a constrained MDP, there may not be a decentralized optimal policy~\cite{altman} Ch 5, we show that for the above video download problem the optimal policy is decentralized when the AP is average power constrained. The clients are coupled through a price for energy $\lambda$ set by the AP. Furthermore, we show that the optimal policy can be obtained by solving a Linear Program in which the number of variables scales linearly with the number of clients $N$. Thirdly, we provide a distributed iterative algorithm to determine the optimal decentralized policy, each step of which involves the clients solving an unconstrained MDP. Since this unconstrained MDP involves minimization of a client's ``local cost" only, each single client's MDP can be solved independently of other clients in a distributed fashion. Each client's optimal policy is shown to be of threshold type, meaning that the policy switches to lower resolution video packets or transmits them at higher power when the playback buffer level drops below certain thresholds. 
When the AP has fewer orthogonal channels $M$ than clients, the AP has to prioritize the clients.
and also choose power and resolution of video packets. We derive an index-based policy that can be viewed as an extension of the Whittle's index policy to the case of Bandit superprocesses~\cite{mahajan_bandit}.

The rest of this paper is organized as follows. We describe previous work in Section~\ref{previousworks}.
We describe the system model in Section~\ref{sec:sd}.
We establish a decentralized optimal randomized solution in Section~\ref{decentralizedoptimalsolution}, and an iterative algorithm to compute optimal policy is proposed in Section~\ref{sec:distributed}. The threshold structure of the optimal solution for each client is shown in Section~\ref{sec:thresh}.

In Section~\ref{sec:qpa} we examine the case when the number of orthogonal channels is less than the number of clients, and the AP has to decide the set of clients for whom to send packets, along with the corresponding resolutions and transmission powers. We determine an index policy. In Section~\ref{sec:simu} we provide the results of simulations and conclude in Section~\ref{sec:con}.
\section{Previous Works} \label{previousworks}
Previous works on video streaming have analyzed relevant trade-offs encountered in optimizing the QoE of video streaming. Higher time spent rebuffering, and hence increased delay before video begins, leads to fewer playback interruptions~\cite{6567080}. The trade-off between outage probability and number of initially buffered packets, i.e initial delay time, is analyzed in~\cite{ParandehGheibi, gliang, egger, a4}, while~\cite{yim} studies the effect of variations in the temporal quality of videos on the global video quality. Reference~\cite{Yuedong} studies the impact of flow level dynamics (flows entering and leaving the system) on the streaming QoE, while~\cite{a5} considers the problem of controlling the rate at which a single client requests data from the server in order to closely match the TCP throughput available to it. However, the model in~\cite{a5} assumes that only a single client is present in the network, ignores inherent system randomness and proposes a heuristic scheme. Reference~\cite{6913491} provides an extensive survey on QoE-related works from human computer interaction and networking domains. 

However, the works listed above do not provide any theoretical guarantees on the QoE properties of the proposed schemes. As an example,~\cite{a5} devises a policy to minimize interruptions for a single client. However, a network-wide deployment of such a policy at each client need not maximize the combined QoE, i.e., a client-by-client optimization need not maximize the overall QoE experienced by the set of all clients~\cite{ho1972team}.


 In order to maximize the cumulative QoE associated with the $N$ clients, a centralized controller that has knowledge of the parameters of each client, and has access to the global system state $\left(\ell_1(t),\ell_2(t),\ldots,\ell_N(t)\right)$ comprised of the instantaneous playback buffer levels $\ell_n(t),n=1,2,\ldots,N$ is essential. However, in this paper we show that the problem admits decentralized policies that have provably optimal QoE guarantees. \par
Another decentralized optimal control problem that arises in a very different context~\cite{singh2016throughput} is to maximize throughput of packets subject to end-to-end deadlines in multi-hop networks. A packet-packet decoupling results there from Lagrangian relaxation.\par
A preliminary announcement of some of these results was presented in the conference paper~\cite{SinKum15Video}. A distributed scheme which performed iterations on the energy price charged by the AP, and client-level policies was shown to converge to an ``optimal" price that maximized the value of the dual function. It was then claimed, without a proof, that a decentralized policy in which each client uses its own convergent policy, is optimal. In the present work, we fill the gap by utilizing an averaging technique to yield an optimal solution for the original constrained Markov decision process. We provide a computationally tractable algorithm and prove that it produces an optimal decentralized policy in which each client randomizes between at most two actions in any given state. Finally, while~\cite{SinKum15Video} considers only the case when the AP is average-power-constrained, the present work extends it to the case when the AP has a ``hard constraint" on the number of packets that can be transmitted. We also provide an ``index policy" which extends the Whittle's index policy~\cite{whittle} to the case of Bandit Superprocesses~\cite{mahajan_bandit}. \section{System Description}\label{sec:sd}
Consider a single server serving video to $N$ clients. 
Each client is connected to the AP through a wireless channel, and video data packets are
streamed through it. 
Time is slotted, and the system evolves over time-slots $t=1,2,\ldots$. The time duration of a slot is equal to the time taken by the AP to attempt a single packet transmission. 
The AP has an average power constraint of $\bar{E}$.

Each client $n$ has a finite playback buffer of size $B_n$ packets, and plays a single packet for a duration of $T_n$ time-slots from it.
After it has finished streaming a packet, it fetches the next packet which is enqueued at the head of its playback buffer and proceeds to stream it. However, if it finds that the playback buffer is empty, then the streaming is interrupted, and we say that an ``outage" has occurred.

The server has multiple copies (called files) of the same video. Each contains
an encoding of the same
portion of the video, differing only with respect to the resolution and the bit rate associated with streaming it.
Having multiples files of differing resolutions is advantageous, since if the playback buffer level of a client is too low approaching
an outage, or the wireless channel is very unreliable, then the AP can switch to transmitting a low resolution file which has only a few bits, thereby increasing probability of successful packet delivery, albeit of a low resolution, but preventing a more negatively perceived outage in video streaming. 
The AP can choose the resolution of the packet to be transmitted to client $n$ from the set $\{1,2,\ldots,R_n\}$. 
We will assume that a lower label corresponds to a finer resolution.

A video quality cost of $\lambda_{r,n}$ units is incurred when client obtains a packet of resolution $r$.
We suppose that $\lambda_{r_1,n}<\lambda_{r_2,n}$ 
for $r_1<r_2$, since the resolution of class $r_1$ packet is better than class $r_2$ packet. 

The AP can choose the packet transmission power for client $n$ from the set $\{e_1,e_2,\ldots,e_{M_n}\}$.
We let $e_1=0$ correspond to the case of no packet transmission, utilizing zero transmission power .

The wireless channels connecting the clients to the AP are unreliable.
We let $p_n(r,e)$ be the probability with which a packet of resolution $r$ transmitted at power $e$ gets is
successfully delivered to client $n$. We suppose that $p_n(r,e)$ is
\begin{itemize}
\item Increasing in $r$ for a fixed value of transmission power $e$ (since a packet with fewer bits and a reduced
resolution
has a higher chance of successful delivery), and is
\item Increasing in $e$ for a fixed value of resolution $r$ (a higher power
leads to higher delivery probability).
\end{itemize}
Denote by $\ell_n(t)$ the amount of play time remaining in client $n$'s playback buffer  at time $t$.
Define the functions  
\begin{align}\label{sf}
&\mathcal{S}_n(\ell_n) := 
\begin{cases}
(\ell_n-1)^{+}+T_n,\mbox{ if } \ell_n \leq B_n-T_n+1,\\
\ell_n-1,\mbox{ if } B_n-T_n+1<\ell_n\leq B_n, 
\end{cases}\\
&\mathcal{F}_n(\ell_n) := (\ell_n-1)^{+}.
\end{align}
Then $\mathcal{S}_n(\ell_n(t))$ and $\mathcal{F}_n(\ell_n(t))$ are the playback buffer values at time $t+1$ that result from a successful and failed packet transmission respectively when the playback buffer level of client $n$ is $\ell_n(t)$ at time $t$. 
To see this, note that if the playback buffer level $\ell_n(t)\leq  B_n-T_n+1$, then $\ell_n(t+1)$ is equal to $(\ell_n(t)-1)^{+}+T_n$ with probability $p_n(r,e)$, and $(\ell_n(t)-1)^{+}$ with probability $1-p_n(r,e)$. But if the playback buffer level $\ell_n(t)>B_n-T_n+1$, then the client cannot accept a new packet because it will lead to an overflow of the playback buffer. Thus if $\ell_n(t)>B_n-T_n+1$, the state of the client at time $t+1$ is $\ell_n(t)-1$ with probability $1$.

Let $e_n(t)$ and $r_n(t)$ denote the transmission power and resolution respectively associated with 
the packet transmission for client $n$ during time-slot $t$. Denote by $u_n(t)=\left(e_n(t),r_n(t)\right)$ the control action chosen for client $n$ during time $t$. The control action for the system is the
vector $u(t)=\{u_n(t)\}_{n=1}^{N}$ that describes the transmission power and video packet resolution utilized for each user $n$.

The state of the system at time $t$ is described by the $N$ tuple $x(t):=\left(\ell_{1}(t),\ell_2(t),\ldots,\ell_N(t)\right)$.
It is a finite-state, finite control, controlled Markov chain. 
We will suppose that $p_n(r,e)>0$ for $e \neq 0$, so that the system can reach the empty state $x=(0,0,\ldots,0)$ under any non-idling policy
after some time steps, so there
is only a single ergodic class under any non-idling stationary randomized feedback policy.


\emph{Quality of Experience:}
The video streaming quality experienced by a client depends upon several factors:
\begin{enumerate}
\item The time that a client spends in a state of outage.
We assume that client $n$ experiences a cost of 1 unit for every time slot that is spent in outage.
\item The number of outage ``periods" experienced by a client.
An outage period is a consecutive period of outage slots beginning with a transition from
the client streaming a packet at time $t-1$ to the client facing an outage at time $t$,
and ending with a transition in the reverse direction. 
Let $\lambda_{o,n}$ denote the cost incurred by client $n$ for every such outage period.
The QoE is affected by the number of outage periods; for example,
if a client is required to face $10$ outage slots, it prefers to experience all of these $10$ outage slots in consecutive time-slots, 
rather than experience it in $5$ batches of $2$ time-slot interruptions each. 
The number of outage periods in the first case is equal to $1$, while that in the second case is equal to $5$. 
We assume that for a fixed number of outages, the clients prefer fewer number of outage periods.
\item Average video quality associated with different resolution types.
Let $\lambda_{r,n}$ denote the cost incurred by client $n$ for every packet of resolution $r$ that it plays out.
\item Temporal variations in resolution, i.e., the number of times the AP switches between the packet video-resolutions.
\end{enumerate}
The costs $\lambda_{o,n}, \left\{ \lambda_{r,n}\right\}_{r=1}^{R_n}$ decide the relative importance that is placed on the different competing objectives, and hence the QoE function can be tuned by varying them. 

For simplicity, we begin by considering the optimization with respect to the factors 1)-3) in the above QoE.

Let $\mathbbm{1}(A)$ be the random variable that 
is the indicator function of the event $A$.
Then $\mathbbm{1}(\ell_n(t)=1, \ell_n(t+1)=0) $ equals $1$ if an outage period begins at time $t$, and is $0$ otherwise. 
So the one-step cost
incurred by the client $n$ at time $t$ is
\begin{align}
&\mathbbm{1}(\ell_n(t)=0)
+ \lambda_{o,n} \mathbbm{1}(\ell_n(t)=1,\ell_n(t+1)=0) \notag \\
&+\lambda_{r,n} \mathbbm{1}(\mbox{Packet of resolution } r \mbox{ received at time }t) . \label{onestepcost}
\end{align}
Noting that the packet is successfully
received
with probability $p_n(e_n(t),r_n(t))$, the expected
cost incurred at time $t$ by an action $u_n(t)=\left(e_n(t),r_n(t)\right)$ is $c_n(\ell_n(t),e_n(t),r_n(t))$, where
\begin{align}
&c_n(\ell_n,e_n,r_n) := \mathbbm{1}(\ell_n=0) +p_n(e_n,r_n) \lambda_r \notag \\
&+ \left(1-p_n(e_n,r_n)\right)
\mathbbm{1}(\ell_n=1)\lambda_o.
\end{align}

Under a stationary randomized policy $\pi$ for the controlled Markov chain, the process $x(t)$ evolves as a finite state, finite control set, controlled Markov process, and the problem of designing the streaming policy that maximizes the cumulative QoE of the clients can be posed as the following \emph{Constrained Markov Decision Process} (CMDP)~\cite{altman},
\begin{align}
&\underset{\pi}{\mbox{Min}} \underset{T \to \infty}{\mbox{ limsup}}\frac{1}{T}\mathbb{E}^\pi \sum_{n=1}^N \sum_{t=0}^{T-1} c_n(\ell_n(t),e_n(t),r_n(t)), \label{costavg} \\
&\mbox{subject to } \underset{T \to \infty}{\mbox{ limsup}}\frac{1}{T}\mathbb{E}^\pi\sum_{n=1}^N\sum_{s=0}^{T-1}e_n(t) \leq \bar{E}\label{pmdp}.
\end{align}
Let $\pi^\star$ be an optimal policy for this CMDP. 

Denote by 
\begin{align}
&\bar{C}^{\pi}_n := \underset{T \to \infty}{\mbox{ limsup}}\frac{1}{T}\mathbb{E}^{\pi}\sum_{t=0}^{T-1} c_n(\ell_n(t),u_n(t))\mbox{ and} \label{indclientcost} \\
&\bar{E}^{\pi}_n := \underset{T \to \infty}{\mbox{ limsup}}\frac{1}{T}\mathbb{E}^{\pi}\sum_{t=0}^{T-1} e_n(t), \label{indclientpower}
\end{align}
the time-average cost and average power consumption incurred by client $n$ under a policy $\pi$, respectively.

\section{Optimality of a decentralized solution} \label{decentralizedoptimalsolution}
\subsection{Existence of a decentralized optimal solution}
We note that for a CMDP with multiple controllers, there generally need not exist a decentralized policy that is optimal, see Ch 5 of~\cite{altman}. Hence, one may suspect that a central controller which makes control choices on the basis of global state $\left(l_1(t),l_2(t),\ldots,l_N(t)\right)$ is required. We now present a key result which shows that for the CMDP (\ref{costavg},\ref{pmdp}), we can in fact restrict to decentralized policies without loss of optimality.

We show that the problem of optimally scheduling the AP's packet streaming to the $N$ clients can be
accomplished by a decentralized policy where each client makes its own decisions independently of other clients.
The only coupling between the clients is through a price per unit energy $\lambda_{e}$
announced by the AP.
Each client $n$ simply chooses its own 
stationary randomized policy $\pi_n$ to minimize the
sum of its video quality and power costs:
\begin{align}\label{scp1}
\underset{{\pi_n}}{\mbox{Min}} \underset{T \to \infty}{\mbox{ limsup}}\frac{1}{T}\mathbb{E}^{\pi}\sum_{t=0}^{T-1} \left(c_n(\ell_n(t),u_n(t) + \lambda_{e} e_n(t) \right).
\end{align}
This policy is fully decentralized in that 
each client $n$ chooses its packet to stream and power at which to be transmitted,
denoted $u_n(t)$, randomly according to
a distribution based only on its own state $\ell_n(t)$
at time $t$, independent of other clients' actions.

We note that for a CMDP with multiple controllers, there generally need not be a decentralized policy that is optimal, see Ch 5 of~\cite{altman}. Hence, one may suspect that implementing an optimal policy for a CMDP requires a central controller which makes control choices on the basis of global state $\left(l_1(t),l_2(t),\ldots,l_N(t)\right)$. We now present a key result which shows that for the CMDP~(\ref{costavg},\ref{pmdp}), we can restrict to decentralized policies without loss of optimality.

\begin{theorem}\label{decentralization}
(i) Suppose $\lambda_e\geq 0$ is a price and, for each client $n$, $\pi^{\star}_n$ is an optimal stationary randomized policy
for the Markov Decision Process (MDP)  (\ref{scp1}) of client $n$, such that either $\lambda_e =0$ or
$\sum_{n=1}^N \bar{E}^{\pi^{\star}_n}_n = \bar{E}$.
Then the combined policy
$\pi^{\star}=\otimes \pi^{\star}_n(\lambda)$,
where the clients independently randomize their actions, is optimal for the overall CMDP (\ref{costavg},\ref{pmdp}).\\
(ii) There do exist such a price $\lambda_e$ and a set of stationary randomized policies $\{\pi^{\star}_n: n=1,2,\ldots,N\}$
satisfying (i). \\
(iii) Moreover, in (ii), each client $n$'s stationary randomized policy $\pi^{\star}_n$ 
can be chosen so that it only randomizes its action at at most
one state $\ell_n \in \{1, 2, \ldots , B_n \}$.
\end{theorem}
\begin{IEEEproof}
(i) A CMDP can be posed as a linear program (LP) in which the 
decision variable is the steady-state measure $\mu_\pi$ induced by a policy $\pi$ on the joint state-action space $(X,U)$~\cite{altman}. 
The infinite horizon average cost~\eqref{costavg} is the dot product between $\mu_{\pi}(\cdot,\cdot)$ and the one step cost function $c(\cdot,\cdot)$.\footnote{The number of variables in the LP is however equal to the cardinality of the joint state-action space.
It therefore increases exponentially with number of clients $N$, and so it is prohibitively intractable to solve
this LP to determine the optimal policy. In Theorem \ref{tractableLP} we therefore provide an alternative tractable
linear complexity solution.}

 Let $\lambda_e$ be the Lagrange multiplier associated with the average-power constraint $\sum_{n=1}^N \bar{E}^{\pi}_n\leq E$. The Lagrangian for the CMDP ~\eqref{costavg}-\eqref{pmdp} is then given by
\begin{align}\label{lagrange}
\mathcal{L}(\pi,\lambda_e) &:= \sum_{n=1}^N \bar{C}^{\pi}_n+ \lambda_{e}\left(\sum_{n=1}^N\bar{E}^{\pi}_n-\bar{E}\right)\notag\\
& = \sum_{n=1}^N \left(\bar{C}^{\pi}_n+ \lambda_{e}\bar{E}^{\pi}_n\right)-\lambda_e\bar{E}.
\end{align}
The dual function is given by,
\begin{align}\label{dual}
D(\lambda_e) := \min_{\pi}\mathcal{L}(\pi,\lambda_e),
\end{align}
and the Dual Problem is,
\begin{align}\label{dualprob}
\max_{\lambda_e\geq 0} D(\lambda_e).
\end{align}
As can be seen in~\eqref{lagrange}, the Lagrangian decomposes into the sum of individual costs $(\bar{C}^{\pi_n}_n+\lambda_e \bar{E}^{\pi_n}_n)$ incurred by each client $n$.
Each client's optimal solution being given by state-action probabilities corresponds to a stationary
randomized policy for its own MDP (\ref{scp1}). If either $\lambda_e =0$, or
$\sum_{n=1}^N \bar{E}^{\pi_n}_n = \bar{E}$ for this set of stationary
randomized policies for the clients, then complementary slackness is satisfied.
So  $\lambda_e$ is optimal for the Dual Problem, and the combination of the stationary randomized policies 
$\pi=\otimes \pi^{\star}_n(\lambda)$ is optimal for the Primal. \\
(ii) To show the existence of such a price $\lambda_e$ and a set of stationary randomized policies $\{\pi^{\star}_n: n=1,2,\ldots,N\}$,
we again start with the linear program.
Let $\pi$ be an optimal stationary randomized policy inducing a
steady-state measure $\mu_\pi$ on the joint state-action space $(X,U)$
that is optimal for this primal LP, and let $\lambda_e$ be the optimal solution of the dual.
Being primal and dual optimal, they satisfy complementary slackness.
That is, either $\lambda_e=0$, or the power control constraint \eqref{pmdp} is satisfied with equality.

However, this stationary randomized policy associates with each complete state
$x=(\ell_1, \ell_2, \ldots , \ell_N)$ a probability distribution for $u=(u_1, u_2, \ldots , u_N)$, while we instead
seek a decentralized policy which has the more stringent property
that each client $n$ can randomize its choice $u_n$ independently
based on its own state $\ell_n$.
Note that for client $n$, $\bar{C}^{\pi}_n$ is the cost attained in \eqref{indclientcost} by this policy $\pi$,
and $\bar{E}^{\pi}_n$ is the power consumed as in \eqref{indclientpower}.\par 
Now consider the following individual CMDP for client $n$, 
\begin{align}
&\mbox{Min }\underset{T \to \infty}{\mbox{ limsup}}\frac{1}{T}\mathbb{E} \sum_{s=0}^{T-1} c_n(\ell_n(t),u_n(t)), \label{clientnCMDPcost}
\end{align}
subject to an individual power constraint
\begin{align}
&\underset{T \to \infty}{\mbox{ limsup}}\frac{1}{T}\mathbb{E} \sum_{s=0}^{T-1} e_n(t) \leq \bar{E}_n.\label{clientnCMDPconstraint}
\end{align}
The LP corresponding to this individual CMDP is feasible since in the overall system the policy $\pi$
attains this power constraint \eqref{clientnCMDPconstraint}.\par 
Let $\hat{\pi}_n$ be an
optimal stationary randomized policy for client $n$ for this individual CMDP
\eqref{clientnCMDPcost}, \eqref{clientnCMDPconstraint}.
The minimum cost \eqref{clientnCMDPcost} attainable is no more than $\bar{C}^{\pi}_n$ since
the policy $\pi$ in the overall system
can indeed attain this cost while satisfying the
power constraint \eqref{clientnCMDPconstraint}. At the same time, the minimal cost
\eqref{clientnCMDPcost} cannot be strictly less
than $\bar{C}^{\pi}_n$, since otherwise if each client uses $\hat{\pi}_n$, the total cost 
of all $N$ clients would be less than that
of $\pi$. Thus, each $\hat{\pi}_n$ attains 
exactly the cost  $\bar{C}^{\pi}_n$ for client $n$ and consumes power $\bar{E}^{\pi}_n$.
Therefore  the combined policy
$\hat{\pi}=\otimes \hat{\pi}_n(\lambda)$ is optimal,
and (ii) follows since
each policy $\hat{\pi}_n$ randomizes independently of the others. \\
(iii) This follows simply since in a CMDP the number of states at which randomization is needed
is at most equal to the number of constraints~\cite{altman}.
Thus each client's policy $\hat{\pi}_n$, that solves the CMDP~(\ref{clientnCMDPcost},\ref{clientnCMDPconstraint}), may be chosen so that it requires randomization in at most one state of
client $n$.
\end{IEEEproof}
%
%
\begin{remark}
Existing results in the literature on CMDPs~\cite{altman} tell us that for the CMDP~(\ref{costavg}-\ref{pmdp}), there is a centralized stationary randomized policy that requires randomization in at most one state. However, this randomization may require the clients to coordinate and require a centralized controller. The above result shows that optimality can be achieved in a decentralized way by each client independently randomizing in just one of its states.
\end{remark}
\subsection{Tractable computation of a decentralized optimal solution}
Now we show that the optimal policies of the clients can be tractably determined.
We show that there is a linear program with $\sum_{n=1}^N B_n M_n R_n = O(N)$ decision variables and 
$1+N+\sum_{n=1}^N B_n M_n R_n = O(N)$ constraints,
that yields the decentralized stationary randomized policies for all the clients.
The number of decision variables therefore grows only linearly in the number of clients.
This should be contrasted with the $\otimes_{n=1}^NB_n M_n R_n$ decision variables in the Linear Program
for the CMDP (\ref{clientnCMDPcost},\ref{clientnCMDPconstraint}), which grows exponentially in the
number of clients.

Below we denote by $\mathcal{S}_n^{-1}$ and $\mathcal{F}_n^{-1}$, the 
set-valued inverse images
of the maps $\mathcal{F}_n$ and $\mathcal{S}_n$ defined in \eqref{sf}.
\begin{theorem}\label{tractableLP}
Consider the following linear programming with decision variables
$\{ \alpha_{\ell ern}: 1 \leq l \leq B_n, 1 \leq e \leq M_n, 1 \leq r \leq R_n, 1 \leq n \leq N \}$:
\begin{align}
&\mbox{Min } \sum_{n=1}^N \sum_{\ell=1}^{B_n} \sum_{e=1}^{M_n} \sum_{k=1}^{R_n} \alpha_{\ell ern} c_n(\ell,e,r) \label{overallobj}
\end{align}
subject to
\begin{align}
& \sum_{e=1}^{M_n}\sum_{r=1}^{R_n} \alpha_{\ell ern} = \sum_{i \in \mathcal{S}_n^{-1}( \ell )} \sum_{j=1}^{M_n} \sum_{k=1}^{R_n} \alpha_{ijkn} p_n(j,k) \notag \\
& \quad + \sum_{i \in \mathcal{F}_n^{-1}( \ell )} \sum_{j=1}^{M_n} \sum_{k=1}^{R_n} \alpha_{i ern} (1-p_n(j,k)), \forall n \label{const1n} \\
&\alpha_{\ell ern} \geq 0, \sum_{\ell=1}^{B_n} \sum_{e=1}^{M_n} \sum_{k=1}^{R_n} \alpha_{\ell ern} = 1, \forall n,  \label{const2n} \\
&\sum_{n=1}^N \sum_{\ell=1}^{B_n} \sum_{e=1}^{M_n} \sum_{k=1}^{R_n} \alpha_{\ell ern} e \leq \bar{E}. \label{constall}
\end{align}
With $\{ \alpha_{\ell ern} \}$ denoting the optimal solution, set 
\begin{align}
&\pi_n^\ell (e,r) := \frac{\alpha_{\ell ern}}{\sum_{j=1}^{M_n} \sum_{k=1}^{R_n} \alpha_{\ell jkn}}.
\end{align}
(In case the denominator above is zero, choose any $\pi_n^\ell (e,r) \geq 0$ satisfying
$\sum_{e=1}^{M_n} \sum_{r=1}^{R_n} \pi_n^\ell (e,r) = 1$). 
Let $\pi_n$ denote the stationary randomized policy for client $n$ that chooses a resolution $r$ and an
energy level $e$ with probability $\pi_n^\ell (e,r)$ whenever its buffer length is $\ell$,
independently of all other clients' actions.
Then the policy $\pi=\otimes \pi_n$ is optimal for the CMDP
\eqref{costavg},\eqref{pmdp}.
\end{theorem}
\begin{IEEEproof}
The constraints \eqref{const1n} and \eqref{const2n} capture the state-action probabilities
for the individual CMDP \eqref{clientnCMDPcost}, \eqref{clientnCMDPconstraint}
for client $n$. The inequality \eqref{constall} constrains the overall power consumed by all clients to \eqref{pmdp}, while
the objective \eqref{overallobj} minimizes the overall cost for the CMDP \eqref{costavg}.
As Theorem \ref{decentralization}.iii shows, these decision variables
yield an optimal policy that results in the minimum cost.
\end{IEEEproof}


\section{Distributed Computation of Optimal Policy}\label{sec:distributed}
Though the results in the previous section yield decentralized policies $\pi=\otimes_{n=1}^{N}\pi_n$, solving the LP~\eqref{overallobj}-\eqref{constall} requires that the parameters $\{c_n(\cdot,\cdot,\cdot),p_n(\cdot)\}_{n=1}^{N}$ be known to a central coordinator (e.g., the AP), which then computes and communicates the optimal policy $\pi_n$ for each user $n$. Next, we devise a distributed iterative scheme which yields an optimal decentralized policy.
\subsection{Distributed computation of optimal price}
First we consider the problem of determining the optimal price $\lambda_e$ in a distributed way.
Consider the Lagrangian~\eqref{lagrange} for the CMDP~\eqref{costavg}-\eqref{pmdp},
\begin{align*}
\mathcal{L}(\pi,\lambda_e) = \sum_{n=1}^N \left(\bar{C}^{\pi}_n+ \lambda_{e}\bar{E}^{\pi}_n\right)-\lambda_e\bar{E}.
\end{align*}
Denote by $\pi_n^{\star}(\lambda_e)$ an optimal stationary randomized solution of the following ``single client MDP" that is parametrized by energy price $\lambda_e$:
\begin{align}\label{eq:mdp_single}
\min_{\pi_n} \left(\bar{C}^{\pi_n}_n+ \lambda_{e}\bar{E}^{\pi_n}_n\right), n=1,2,\ldots,N.
\end{align}
The MDP~\eqref{eq:mdp_single} can be posed as the following LP,
\begin{align}
\mbox{Min } \sum_{\ell=1}^{B_n} \sum_{e=1}^{M_n} \sum_{k=1}^{R_n} \alpha_{\ell er} c_n(\ell,e,r)+\lambda_e e \label{overallobj_single}
\end{align}
subject to
\begin{align}
& \alpha_{\ell ern} = \sum_{i \in \mathcal{S}_n^{-1}( \ell )} \sum_{j=1}^{M_n} \sum_{k=1}^{R_n} \alpha_{ijkn} p_n(j,k) \notag \\
& \quad + \sum_{i \in \mathcal{F}_n^{-1}( \ell )} \sum_{j=1}^{M_n} \sum_{k=1}^{R_n} \alpha_{\ell ern} (1-p_n(j,k)), \forall n \label{const1n_single} \\
&\alpha_{\ell ern} \geq 0, \sum_{\ell=1}^{B_n} \sum_{e=1}^{M_n} \sum_{k=1}^{R_n} \alpha_{\ell ern} = 1, \label{const2n_single}.
\end{align}
If $\alpha_n(\lambda_e):=\{ \alpha_{\ell ern} \}_{l \in [1,B_n], e \in [1,M_n], r \in [1,R_n]}$\footnote{For two integeres $x,y$ with $y>x$, we let $[x,y]:=\left\{x,x+1,\ldots,y\right\}$.} denotes an optimal solution to the above LP, then,
\begin{align}\label{eq:opt_action}
\pi_n^{\star,\ell} (e,r) := \frac{\alpha_{\ell ern}}{\sum_{j=1}^{M_n} \sum_{k=1}^{R_n} \alpha_{\ell jkn}}
\end{align}
is optimal for~\eqref{eq:mdp_single}.
Clearly, the policy $\otimes_{i=1}^{N} \pi^{\star}_n(\lambda_e)$ maximizes the Lagrangian $\mathcal{L}(\cdot,\lambda_e) $, and hence corresponds to evaluating the dual function $D(\lambda_e)$, i.e. 
\begin{align*}
D(\lambda_e) = \sum_{n=1}^N \left(\bar{C}^{\pi^{\star}_n(\lambda_e)}_n+ \lambda_{e}\bar{E}^{\pi^{\star}_n(\lambda_e)}_n\right)-\lambda_e\bar{E}.
\end{align*}
Thus, $\frac{\partial D(\lambda_e)}{\partial \lambda_e} = \sum_{n=1}^N \bar{E}^{\pi^{\star}_n(\lambda_e)}_n-\bar{E}$.
Since the dual function is concave, price tatonnement iterations 
$\lambda_e^{k} = \lambda_e^{k} + \frac{1}{k}\left(\sum_{i=1}^{N} \bar{E}^{\pi^{\star}_n(\lambda_e^{k})}_n - \bar{E}\right)$,
correspond to gradient ascent algorithm applied to solve the dual problem
\begin{align}\label{eq:dual}
\max_{\lambda_e\geq 0} D(\lambda_e),
\end{align}
and hence converge to $\lambda^{\star}_e$ which maximizes the dual function $D(\lambda_e)$~\cite{Ber87}.

The iterations involving $\lambda^{(k)}_e$ can be performed in a distributed way as follows. The AP declares the price $\lambda^{(k)}_e$ to all the clients. Then, each client $n$ can solve for the policy $\pi_n^{\star}(\lambda^{(k)}_e)$ and $\alpha_n(\lambda^{(k)}_e)$ in a distributed way since solving the LP~\eqref{overallobj_single}-\eqref{const2n_single} requires the knowledge of the parameters of client $n$ only. Next, the clients communicate their energy utilizations $\{\bar{E}^{\pi^{\star}_n(\lambda_e^{(k)})}_n\}_{n=1}^{N}$ to the AP, which are then used by the AP to update the price $\lambda^{(k)}_e$. 
\subsection{Distributed computation of optimal policies}
It might occur that the policy $\otimes_{i=1}^{N} \pi^{\star}_n(\lambda^{\star}_e)$ is optimal for the CMDP~(\ref{costavg}-\ref{pmdp}). However, this may not always be the case since even though the policy $\otimes_{i=1}^{N} \pi^{\star}_n(\lambda^{\star}_e)$ optimizes the Lagrangian $\sum_{n=1}^N \left(\bar{C}^{\pi}_n+ \lambda^{\star}_{e}\bar{E}^{\pi}_n\right)$, and hence from strong duality also yields the optimal cost, it may not satisfy the energy constraints $\sum_{n=1}^{N}\bar{E}^{\pi^{\star}_n(\lambda^{\star}_e)}_n \leq \bar{E}$. For a general convex optimization problem, given a dual solution, it is not straightforward to recover an optimal primal solution unless the dual function $D(\lambda_e)$ is differentiable at $\lambda^{\star}_e$~\cite{gustavsson2015primal}. In our set-up the dual function $D(\lambda_e)$ is piecewise linear since there are only a finite number of stationary (non-randomized) policies, and each such policy is optimal for the cost function $\left(\bar{C}^{\pi_n}_n+ \lambda_{e}\bar{E}^{\pi_n}_n\right)$ for values of $\lambda_e$ lying within a closed interval of the real line.

To address this problem, one can use the averaging procedure of~\cite{gustavsson2015primal} which shows that a weighted convex combination of the iterates $\left\{\alpha_n(\lambda_e^{(k)})\right\}_{n=1}^{N}, k=1,2,\ldots$ does converge to a solution of the original constrained problem. 
\begin{figure}[!t]
	\centering
	\includegraphics[width=0.5\textwidth]{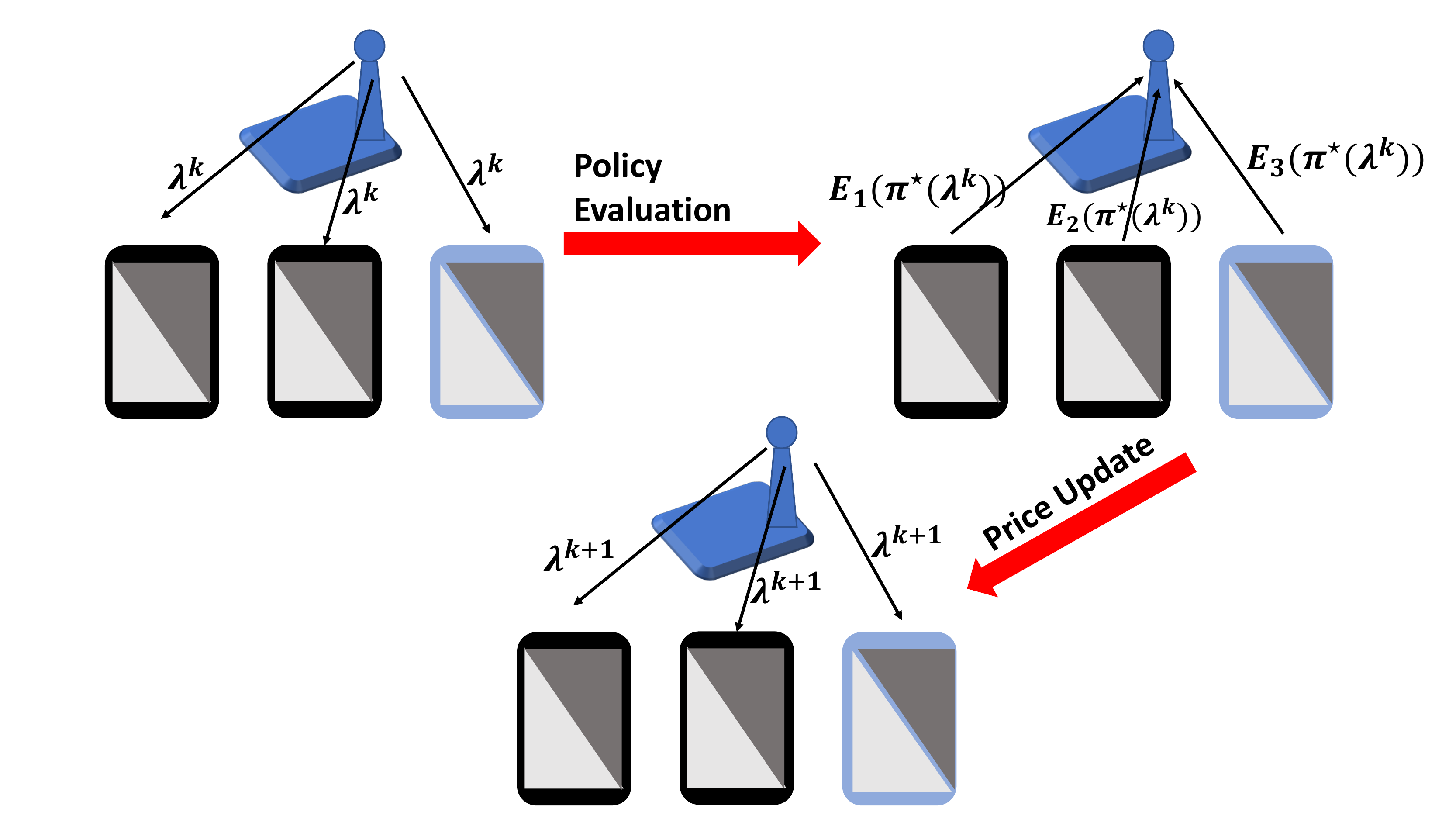}
	\caption{Decentralized iterations involving policy evaluations (at clients) followed by price updates (at AP) with appropriate step-sizes converge to the optimal price $\lambda^\star$, and, after averaging, converge to the optimal policy $\otimes \pi^\star(\lambda^\star_e)$.}
	\label{fig:iterations}
\end{figure}

\begin{theorem}\label{th:weighted_avg}
Consider the following iterative procedure,
\begin{align}
\bar{\alpha}^{(k)} &= \sum_{s=1}^{k} \gamma^{(s,k)} \otimes_{n=1}^{N} \alpha^{(s)}_n, \label{policy_update}\\
\lambda_e^{(k+1)} &= \lambda_e^{k} +\beta^{(k)}\left(\sum_{i=1}^{N} \bar{E}^{\pi^{(k)}_n }_n - \bar{E}\right),\label{price_update} k=1,2,\ldots,
\end{align}
where~\eqref{policy_update} denotes the elementwise vector addition of the vectors, step-sizes $\beta^{(k)} = 1\slash k$, $\gamma^{(s,k)}: = \beta^{(s)} \slash \sum_{t=1}^{k} \beta^{(t)}$. Then, $\bar{\alpha}^{(k)}  \to\alpha^{\star}$, which solves~\eqref{costavg}-\eqref{pmdp}. 
\end{theorem}
\begin{IEEEproof}
The result follows from Corollary 5 of~\cite{anstreicher2009two}. 
\end{IEEEproof}\par 
It should be noted that the iterations can be implemented in a distributed manner.
\section{Computation of Structured Distributed Policies}\label{sec:structure}
Let $\Pi_{n,det}$ denote the class of stationary deterministic (``pure") policies for client $n$. The number of such policies is equal to $\left(R_nM_n\right)^{B_n}$. We now show that a simple modification to the iterations~(\ref{policy_update},~\ref{price_update})
yields a decentralized policy under which the occupation measures of the client-level optimal policies produced by the algorithm $\pi^{\star}_n,n=1,2,\ldots,N$ are a convex combination of the occupation measures of at most two pure policies from $\Pi_{n,det}$. Consequently, the resulting policy randomizes amongst at most two actions in each value of the buffer level $\ell$. Such a result is important because even though Theorem~\ref{decentralization} shows that there is an optimal policy under which each client utilizes randomization in at most one state, it does not provide a tractable algorithm to produce such a policy.
\begin{lemma}\label{lemma:dual_pwl}
Define the functions
\begin{align}\label{eq:hatd}
\hat{D}_{n}(\lambda_e):= \min_{\pi_n} \left[\bar{C}^{\pi_n}_n+ \lambda_{e}\bar{E}^{\pi_n}_n\right], n=1,2,\ldots,N,
\end{align}
where the minimization is over the class of stationary randomized policies. The functions $\hat{D}_n(\lambda_e),n=1,2,\ldots,N$ are piecewise linear since they are minima of a finite number of linear functions, one corresponding to each of the finite number of policies. Let $\mathcal{I}^{n}_{1},\mathcal{I}^{n}_{2},\ldots,\mathcal{I}^{n}_{K_n}$ be the corresponding minimal partition into finitely many contiguous intervals, with a common minimizer in each interval. Let $\Pi_{n,det,\mathcal{I}^{n}_{k}}$ be the set of deterministic stationary policies that minimize the cost $\bar{C}^{\pi_n}_n+ \lambda_{e}\bar{E}^{\pi_n}_n$ for values $\lambda \in \mathcal{I}^{n}_{k}$. The slope of $\hat{D}_{n}(\lambda_e)$ within the interval $\mathcal{I}^{n}_{k}$ is equal to the average energy utilization $\bar{E}^{\pi_n}_n$ of policies $\pi_n \in \Pi_{n,det,\mathcal{I}^{n}_{k}}$, and moreover $\Pi_{n,det,\mathcal{I}^{n}_{i}} \cap \Pi_{n,det,\mathcal{I}^{n}_{j}}=\phi$ if $i\neq j$  by the minimality of the partition. We take the closure of each interval, and, abusing notaiton, still denote them by $\mathcal{I}_k^n$. A single policy is still a minimizer throughout the closed interval. At the common boundary point corresponding to the intersection of two closed intervals, two policies are optimal.
\end{lemma}
\emph{Iterative Algorithm with Memory}:
Consider the iterative procedure~(\ref{policy_update},~\ref{price_update}) of Theorem~\ref{th:weighted_avg}, and modify it slightly as follows. The clients maintain a set of policies $\mathcal{U}^{(k)}_n$ containing at most two policies. During iteration $k=1,2,\ldots$, the clients consider their individual MDP \begin{align}\label{eq:client_det_mdp}
\min_{\pi_n \in \Pi_{n,det}} \left[\bar{C}^{\pi_n}_n+ \lambda^{(k)}_{e}\bar{E}^{\pi_n}_n\right].
\end{align} 
They verify whether a policy from the set $\mathcal{U}^{(k)}_n$ solves the MDP~\eqref{eq:client_det_mdp}. If a policy in the set $\mathcal{U}^{(k)}_n$ solves this MDP, then it sets $\alpha^{(k)}_n$ equal to its occupation measure. Otherwise, it includes in the set $\mathcal{U}^{(k)}_n$ the new deterministic policy which solves the MDP, and sets $\alpha^{(k)}_n$ equal to its occupation measure. If the size of the obtained set is greater than $2$, then remove the ``older" 
policy which is not optimal for~\eqref{eq:client_det_mdp} from $\mathcal{U}^{(k)}_n$. Also denote by $\pi^{(k)}_n$ the policy that was added to $\mathcal{U}^{(k)}_n$, or which was already present in $\mathcal{U}^{(k)}_n$ and optimized the MDP.
\begin{lemma}\label{lemma:2}
For the modified iterative algorithm, let $\lambda^{\star}_e$ denote the limiting value of price. Let $\epsilon>0$ be smaller than the minimum of the length of the intervals $\left\{\mathcal{I}^{n}_{j}\right\}_{j=1}^{K_n}, n=1,2,\ldots,N$. The sets $\mathcal{U}^{(k)}_{n},n=1,2,\ldots,N$ converge as $k\to\infty$ to a set~$\mathcal{U}_n^{\infty}$ containing at most two policies. There is a $k_0$ such that $\mathcal{U}_n^{(k)} = \mathcal{U}_n^{\infty},~\forall k>k_0$.
Also,
\begin{enumerate}
\item Either, $\lambda^{\star}_{e}$ is the common boundary point of two contiguous intervals $\mathcal{I}^{n}_{j},\mathcal{I}^{n}_{j+1}$,
\item Or $\lambda^{\star}_{e}$ belongs to the interior of an interval $\mathcal{I}^{n}_{j}$.
\end{enumerate}
In either of the two cases, we have that $\forall \lambda_e \in \left(\lambda^{\star}_{e}-\epsilon,\lambda^{\star}_{e}+\epsilon\right), \exists \pi_n \in  \mathcal{U}^{\infty}_{n}$ such that $\pi_n$ is optimal for MDP~\eqref{eq:client_det_mdp}.
\end{lemma}
\begin{IEEEproof}
We only consider the first case when $\lambda^{\star}_{e}$ is the common boundary point of $\mathcal{I}^{n}_{j},\mathcal{I}^{n}_{j+1}$. Noting that $\lambda_e^{(k)} \to \lambda_e^{\star}$, let $k_1$ be such that $\lambda^{(k)}_e \in \left(\lambda^{\star}_{e}-\epsilon,\lambda^{\star}_{e}+\epsilon\right)~\forall k>k_1$. Without loss of generality, let $\lambda^{(k_1)}_e \in \mathcal{I}^{n}_{j}$. 
Let $k_2>k_1$ be the first iteration after $k_1$ such that $\lambda^{(k)}_e \in \mathcal{I}^{n}_{j+1}\setminus \partial \mathcal{I}^{n}_{j+1}$\footnote{The case when $k_2 = \infty$ is similar and not considered here.}. It follows from construction of the algorithm that the policy $\pi^{(k_1)}_{n}$ was played during iterations $k_1,k_1 +1,\ldots,k_2 - 1$. Since in any iteration, an older policy is deleted from the set $\mathcal{U}^{(k)}_n$, the set $\mathcal{U}^{(k_2 -1)}_{n}$ contains $\pi^{(k_1)}_{n}$. 

From Lemma~\ref{lemma:dual_pwl} we have that $\pi^{(k_1)}_{n}$ is not optimal for $\lambda_e \in  \mathcal{I}^{n}_{j+1}\setminus \partial \mathcal{I}^{n}_{j+1}$, and hence we have $\pi^{(k_1)}_{n} \neq \pi^{(k_2)}_{n}$. Thus, the set $\mathcal{U}^{(k_2)}_{n}$ consists of two distinct policies $\pi^{(k_1)}_{n},\pi^{(k_2)}_{n}$. It follows from the construction of the algorithm that for $k>k_2$, $\pi^{(k_1)}_{n}$ is played if $\lambda^{(k)}_e \in \mathcal{I}^{n}_{j}$, while $\pi^{(k_2)}_{n}$ is played if $\lambda^{(k)}_e \in \mathcal{I}^{n}_{j+1}$\footnote{Ties are broken according to some fixed rule.}, so that the set $\mathcal{U}^{(k)}_{n} = \left\{ \pi^{(k_1)}_{n},\pi^{(k_2)}_{n} \right\},\forall  k>k_2$. Thus, $\mathcal{U}^{\infty}_{n} = \left\{ \pi^{k_1}_{n},\pi^{k_2}_{n} \right\}$. These policies clearly solve~\eqref{eq:client_det_mdp} when price $\lambda_e$ lies in their corresponding intervals.
\end{IEEEproof}
\begin{algorithm}
\caption{Distributed Algorithm to Yield Two Critical Pure Polices Providing an Optimal Randomization }
\begin{algorithmic}\label{algo:1}
\STATE $k=0,~\mathcal{U}^{(0)}_n = \emptyset, \forall n=1,2,\ldots,N$. Set $\lambda^{(0)}_e$ to an arbitrary value.
 \STATE \REPEAT
  \STATE 1.) Client $n$: Verifies whether a policy in $\mathcal{U}^{(k)}_n$ solves~\eqref{eq:client_det_mdp}. If so, set $\pi^{(k)}_n$ equal to it and set  $\mathcal{U}^{(k+1)}_n = \mathcal{U}^{(k)}_n$. Otherwise solve~\eqref{eq:client_det_mdp} to obtain $\pi^{(k)}_{n}$ and update $\mathcal{U}^{(k)}_n$ as follows. If $|\mathcal{U}^{(k)}_n|<2$ then update $\mathcal{U}^{(k+1)}_n = \mathcal{U}^{(k)}_n \cup \left\{\pi^{(k)}_{n} \right\}$. However, if $|\mathcal{U}^{(k)}_n|=2$, then set $\mathcal{U}^{(k+1)}_n = \left\{\mathcal{U}^{(k)}_n \setminus \{ a\} \right\}\cup \left\{\pi^{(k)}_{n} \right\}$, where $a$ is the policy in set $\mathcal{U}^{(k)}_n$ that is ``older". Communicate the average energy consumption $\bar{E}^{\pi^{(k)}_n}_n$ and the occupation measure $\alpha^{(k)}_n$ of the policy $\pi^{(k)}_{n}$ to the AP.
  to the AP. 
 \STATE 2.) AP:  Perform the updates for the joint occupation measure~\eqref{policy_update} and the price~\eqref{price_update}.
 \STATE 3.) $k\to k+1$
\UNTIL{$\lambda^{(k)}_e \to \lambda^{\star}_e,~\bar{\alpha}^{(k)}  \to \alpha^{\star}$.}
\STATE  
 \end{algorithmic} 
 \end{algorithm}
We now show the structure of the optimal distributed policy that results from the above algorithm. In the discussion below, for a deterministic policy $\pi$, with occupation measure $\alpha$, we will occasionally use $\alpha$ to denote the policy.
\begin{theorem}\label{th:binary_combine}
Consider the following linear program with decision variables restricted to $\{ \alpha_{\ell ern}: 1 \leq l \leq B_n, (e,r) \in  \left\{\pi^{\infty}_{n,1}(\ell)  ,\pi^{\infty}_{n,2}(\ell)\right\}  , 1 \leq n \leq N \}$, i.e., restricted to only those state-action pairs that can result only from the policies produced by Algorithm 1:
\begin{align}
&\mbox{Min } \sum_{n=1}^N \sum_{\ell=1}^{B_n} \sum_{(e,r) \in \left\{\pi^{\infty}_{n,1}(\ell)  ,\pi^{\infty}_{n,2}(\ell)\right\}  }\alpha_{\ell ern} c_n(\ell,e,r) \label{overallobj_sparse}
\end{align}
subject to
\begin{align}
& \sum_{(e,r) \in \left\{\pi^{\infty}_{n,1}(\ell)  ,\pi^{\infty}_{n,2}(\ell)\right\}  } \alpha_{\ell ern} = \\
&\sum_{i \in \mathcal{S}_n^{-1}( \ell )} \sum_{(j,k) \in \left\{\pi^{\infty}_{n,1}(i)  ,\pi^{\infty}_{n,2}(i)\right\}  }\alpha_{ijkn} p_n(j,k) \notag \\
& \quad + \sum_{i \in \mathcal{F}_n^{-1}( \ell )}\sum_{(j,k) \in \left\{\pi^{\infty}_{n,1}(i)  ,\pi^{\infty}_{n,2}(i)\right\}  }\alpha_{i ern} (1-p_n(j,k)), \forall n \label{const1n_sparse} \\
&\alpha_{\ell ern} \geq 0, \sum_{\ell=1}^{B_n} \sum_{(e,r) \in  \left\{\pi^{\infty}_{n,1}(\ell)  ,\pi^{\infty}_{n,2}(\ell)\right\}  }\alpha_{\ell ern} = 1, \forall n,  \label{const2n_sparse} \\
&\sum_{n=1}^N \sum_{\ell=1}^{B_n} \sum_{(e,r) \in \left\{\pi^{\infty}_{n,1}(\ell)  ,\pi^{\infty}_{n,2}(\ell)\right\}  }  \alpha_{\ell ern} e \leq \bar{E}. \label{constall_sparse}
\end{align}
\end{theorem}
\begin{IEEEproof}
First we establish the properties of Algorithm~\ref{algo:1}. The price iterations correspond to the sub-gradient ascent method applied to solve the dual problem~\eqref{eq:dual}. With $\beta^{(k)}=1\slash k$, we have $\sum_{k=1}^{\infty} \beta^{(k)} = \infty,\sum_{k=1}^{\infty} \left(\beta^{(k)}\right)^2 <\infty$, and hence it follows from Theorem 2 of~\cite{anstreicher2009two} that in Algorithm~\ref{algo:1} $\lambda^{(k)}_e\to\lambda^{\star}$, where $\lambda^{\star}$ is optimal for the dual problem~\eqref{eq:dual}. 

Consider now the average the of occupation measures formed from the iterates of the algorithm, as in Theorem 4 of~\cite{anstreicher2009two}. Any accumulation point $\alpha = \left\{\alpha(\ell_1, u_1, \ell_2, u_2, \ldots,\ell_N, u_N)\right\}: \forall$ states $\ell_i$ and action $u_i$ of client $i$ is an optimal occupation measure, i.e., it is an optimal solution of the primal CMDP ~(\ref{costavg}-\ref{pmdp}).

Now we show that any accumulation point $\alpha^{\star}$ has the property that it is in the convex hull of products of occupation measures produced by convex combinations of $\pi_{n,1}^\infty$ and $\pi_{n,2}^\infty$ for each client $n$, which set we will denote by $\otimes_{n=1}^{N} \mathcal{U}^{\infty}_n$. For each deterministic policy $\pi$ for the overall system composed of $N$ clients define 
\begin{align}
\mathcal{T}_{k,\pi} : = \left\{ t\in\mathbb{N} |~t\leq k, \alpha^{(t)}  = \alpha_{\pi}\right\},
\end{align}
i.e. those iteration indices until iteration number $k$ during which the optimal deterministic policy for iteration $k$ under price $\lambda^{(k)}_e$ was $\pi$. Since there are only finitely many deterministic policies for the overall system, we have that the occupation measure in~\eqref{policy_update} can equivalently be written as 
\begin{align}
\bar{\alpha}^{(k)} = \sum_{\pi \in \Pi_{det}} \alpha_{\pi} \frac{\sum_{s\in \mathcal{T}_{k,\pi}}\beta^{(s)}}{\sum_{t=1}^{k} \beta^{(t)}}.
\end{align} 
 The weights $\frac{\sum_{s\in \mathcal{T}_{k,\pi}}\beta^{(s)}}{\sum_{t=1}^{k} \beta^{(t)}}$ are clearly non-negative and sum to $1$.
Thus, $\bar{\alpha}^{(k)} $ is a convex combination of the occupation measures $\left\{\alpha_{\pi}\right\}_{\pi\in \Pi_{det}}$. Therefore, it remains only to show that asymptotically, only the weights associated with the occupation measures in the set $\otimes_{n=1}^{N} \mathcal{U}^{\infty}_n$ are non-zero. It follows from Lemma~\ref{lemma:2} that for $k>k_0$, each client $n$ plays a policy from the set $\mathcal{U}^{\infty}_n$. Hence, for $k>k_0$, we have that the measure $\alpha^{(k)}\in \otimes_{n=1}^{N} \mathcal{U}^{\infty}_n$. Thus, for deterministic policies $\pi\notin \otimes_{n=1}^{N} \mathcal{U}^{\infty}_n$, the quantity $\sum_{s\in \mathcal{T}_{k,\pi}}\beta^{(s)}$ remains bounded as $k\to\infty$. Since $\sum_{t=1}^{\infty} \beta^{(t)} =\infty$, we have that $\lim_{k\to\infty} \sum_{s\in \mathcal{T}_{k,\pi}}\beta^{(s)} \slash  \sum_{t=1}^{k} \beta^{(t)}  = 0$ for $\pi\notin \otimes_{n=1}^{N} \mathcal{U}^{\infty}_n$.

Since the following statement holds true for all $\alpha \in \otimes_{n=1}^{N} \mathcal{U}^{\infty}_n$,
\begin{align}
&\alpha(\ell_1,u_1,\ell_2,u_2,\ldots,\ell_N,u_N)  = 0 \mbox{ if }\notag\\
 &\exists n\mbox{ s.t. }, u_n \notin  \left\{\pi^{\infty}_{n,1}(\ell_n)  ,\pi^{\infty}_{n,2}(\ell_n)\right\},
\end{align}
and since any accumulation point $\alpha^{\star}$ belongs to the convex hull of the set $\otimes_{n=1}^{N} \mathcal{U}^{\infty}_n$, we have that 
\begin{align}\label{eq:alpha_star_prop}
&\alpha^{\star}(\ell_1,u_1,\ell_2,u_2,\ldots,\ell_N,u_N)  = 0 \mbox{ if }\notag\\
 &\exists n\mbox{ s.t. }, u_n \notin  \left\{\pi^{\infty}_{n,1}(\ell_n)  ,\pi^{\infty}_{n,2}(\ell_n)\right\}.
\end{align}

It remains to show that occupation measure $\hat{\alpha}^{\star}$ produced by the LP~\eqref{overallobj_sparse}-\eqref{constall_sparse} yields the desired solution. The stationary randomized policy produced by $\alpha^{\star}$ associates with each state for the overall system
$x=(\ell_1, \ell_2, \ldots , \ell_N)$ a probability distribution for $u=(u_1, u_2, \ldots , u_N)$. Let $\bar{E}^{\alpha^{\star}}_n,\bar{C}^{\alpha^{\star}}_n$ be the optimal power consumption and cost incurred respectively by client $n$ under the policy $\alpha^{\star}$. Now consider the following individual CMDP for client $n$, 
\begin{align}
&\mbox{Min }\underset{T \to \infty}{\mbox{ limsup}}\frac{1}{T}\mathbb{E} \sum_{t=0}^{T-1} c_n(\ell_n(t),u_n(t)), \label{clientnCMDPcost_sparse}
\end{align}
subject to an individual power constraint
\begin{align}
&\underset{T \to \infty}{\mbox{ limsup}}\frac{1}{T}\mathbb{E} \sum_{t=0}^{T-1} e_n(t) \leq \bar{E}^{\alpha^{\star}}_n\label{clientnCMDPconstraint_sparse},
\end{align}
and constraints on state-action frequencies
\begin{align}
\underset{T \to \infty}{\mbox{ limsup}}\frac{1}{T}\mathbb{E} \sum_{t=0}^{T-1} \mathbbm{1} \left( (e_n(t),r_n(t)) \notin \right. \notag \\
 \left. \qquad \left\{\pi^{\infty}_{n,1}(\ell_n(t))  ,\pi^{\infty}_{n,2}(\ell_n(t))\right\}   \right) =0.\label{clientn_saf_sparse}
\end{align}

The LP corresponding to this individual CMDP is feasible since in the overall system the policy  $\alpha^{\star}$ satisfies the energy constraint~\eqref{clientnCMDPconstraint_sparse}, and it follows from~\eqref{eq:alpha_star_prop} that it also satisfies the constraints~\eqref{clientn_saf_sparse}.

Let $\alpha^{\star}_n$ be an
optimal stationary randomized policy for client $n$ for this individual CMDP~\eqref{clientnCMDPcost_sparse}-\eqref{clientn_saf_sparse}.
The minimum cost~\eqref{clientnCMDPcost_sparse} is clearly upper-bounded by $\bar{C}^{\alpha^{\star}}_n$ since in the overall system $\alpha^{\star}$ can indeed attain this cost while satisfying the constraints~\eqref{clientnCMDPconstraint_sparse},\eqref{clientn_saf_sparse}. At the same time, the minimal cost \eqref{clientnCMDPcost_sparse} cannot be strictly less
than $\bar{C}^{\alpha^{\star}}_n$, since otherwise if each client uses $\alpha^{\star}_n$, the total cost  of all $N$ clients would be less than that of $\alpha^{\star}$. Thus, each $\alpha^{\star}_n$ attains exactly the cost  $\bar{C}^{\alpha^{\star}}_n$  for client $n$ and consumes power $\bar{E}^{\alpha^{\star}}_n$.
Therefore the decentralized policy $\otimes_{n=1}^{N} \alpha^{\star}_n$ for the combined system is also optimal and can be obtained by solving the LP~\eqref{overallobj_sparse}-\eqref{constall_sparse}.
\end{IEEEproof}
We notice that solving the LP~(\ref{overallobj_sparse}-\ref{constall_sparse}) does not require knowledge of the limiting occupation measure $\alpha^{\star}$. Solving the LP only requires the sets $\mathcal{U}^{\infty}_n,n=1,2,\ldots,N$ as input. Thus, one could forego the iterations~\eqref{policy_update} in order to yield the much simpler Algorithm~\ref{algo:2} described below.
\begin{corollary}
Algorithm~\ref{algo:2} yields an optimal decentralized policy that is structured, i.e, for each of the $N$ clients, the resulting policy randomizes amongst at most two actions in each value of the buffer level $\ell_n$.
\end{corollary}

\begin{algorithm}
\caption{Algorithm to Yield Decentralized Structured Optimal Policy}
\begin{algorithmic}\label{algo:2}
\STATE $k=0,~\mathcal{U}^{(0)}_n = \emptyset, \forall n=1,2,\ldots,N$. Set $\lambda^{(0)}_e$ to an arbitrary value.
 \STATE \REPEAT
 \STATE 1.) Client $n$: Verifies whether a policy in $\mathcal{U}^{(k)}_n$ solves~\eqref{eq:client_det_mdp}. If so, set $\pi^{(k)}_n$ equal to it and set  $\mathcal{U}^{(k+1)}_n = \mathcal{U}^{(k)}_n$. Otherwise solve~\eqref{eq:client_det_mdp} to obtain $\pi^{(k)}_{n}$ and update $\mathcal{U}^{(k)}_n$ as follows. If $|\mathcal{U}^{(k)}_n|<2$ then update $\mathcal{U}^{(k+1)}_n = \mathcal{U}^{(k)}_n \cup \left\{\pi^{(k)}_{n} \right\}$. However, if $|\mathcal{U}^{(k)}_n|=2$, then set $\mathcal{U}^{(k+1)}_n = \left\{\mathcal{U}^{(k)}_n \setminus \{ a\} \right\}\cup \left\{\pi^{(k)}_{n} \right\}$, where $a$ is the policy in set $\mathcal{U}^{(k)}_n$ that is ``older". Communicate the average energy consumption $\bar{E}^{\pi^{(k)}_n}_n$
 to the AP. 
 \STATE 2.) AP:  Perform the updates for the price~\eqref{price_update}.
 \STATE 3.) $k\to k+1$
\UNTIL{$\lambda^{(k)}_e \to \lambda^{\star}_e$.}
\STATE  The clients communicate their $\mathcal{U}^{\infty}_n,n=1,2,\ldots,N$ to the AP. AP computes the optimal structured policy by solving the LP~(\ref{overallobj_sparse}-\ref{constall_sparse}).
 \end{algorithmic} 
 \end{algorithm}

\section{Threshold Structure of Optimal Policy}\label{sec:thresh}
\subsection{Structure of the Optimal Policy for the Single Client MDP}\label{sec:thresh_single}
We now show that the single client MDPs~\eqref{eq:mdp_single} have a monotonicity structure. Thus, any optimal policy for~\eqref{eq:mdp_single} is necessarily monotonic, i.e., it employs actions $u$ which have a higher probability of successful transmission $p(u)$ as the buffer level $\ell$ decreases. 

We actually show a somewhat stronger result, that there is an optimal policy that has a simple threshold structure with respect to the buffer level. Such a policy has buffer level thresholds at which it switches to an action $u$ that has a strictly higher probability $p(u)$ of successfully delivering the packet. 

In this and the following section, since we will only be concerned with optimizing the cost of a single client $n$, we will omit the subscript $n$ associated with variables. We begin with some definitions.
\begin{definition}[Monotone Policy]\label{def:monotone}
 A policy $\pi$ is said to be monotonic if it satisfies the following condition: If the policy $\pi$ chooses an action $u_1 = \left(e_1,r_1\right)$ when
 its buffer level, i.e., its state,  is $\ell+1$, and it chooses an action $u_2 = \left(e_2,r_2\right)$ in state $\ell$, then $p(u_2)\geq p(u_1)$.
\end{definition}

\begin{definition}[Threshold-type Policy] \label{def:threshold}
A policy $\pi$ is of threshold-type if it satisfies the following condition:
If the policy $\pi$ chooses an action $u_1 = \left(e_1,r_1\right)$ when
its buffer level, i.e., its state,  is $\ell+1$, and it chooses an action $u_2 = \left(e_2,r_2\right)$ in state $\ell$, then either $u_1=u_2$ or $p(u_2)>p(u_1)$.
\end{definition}
We note that a threshold-type policy is necessarily monotonic, while a monotonic policy may not be of threshold-type.

For a single client, consider the minimum value of the total cost incurred over a horizon of $s\geq 1$ time-slots,
\begin{align}
& V_s(\ell) := \min_{\pi} \mathbb{E}_{\ell}\sum_{t=0}^{s}\left[c(\ell(t),u(t))+\lambda_e e(t) \right],\label{disc}
\end{align} 
with the subscript in $\mathbb{E}_{\ell}$ denoting that the initial state is $\ell$.
The Dynamic Programming backward recursion is
\begin{align}
V_{s}(\ell)  & = \min_{(e,r)} \left\{\mathbbm{1}(\ell=0)+ \lambda_e e+p(e,r)\left[\lambda_r+ V_{s-1}(\mathcal{S}(\ell)) \right]\right.\notag\\
&\quad \quad  \quad \left.+\left(1-p(e,r)\right)\left[\mathbbm{1}(\ell=1)\lambda_o+ V_{s-1}(\mathcal{F}(\ell))  \right]\right\}. \notag
\end{align}
This can be rewritten as
\begin{align}
V_{s}(\ell)  & =\mathbbm{1}(\ell=0) + \mathbbm{1}(\ell=1)\lambda_o+ V_{s-1}(\mathcal{F}(\ell)) \notag\\
&\quad \quad  \quad + \min_{u=\left(e,r\right)}  \{\hat{c}(u)-p(u) \mathcal{D}_{s-1}(\ell)\}, \label{eq:1} \end{align}
where
\begin{align}
&\hat{c}(u):=\lambda_e e + p(e,r)\lambda_r, \label{cost}
\end{align}
is the one-step augmented cost of choosing $u=(e,r)$, and
\begin{align}
\mathcal{D}_{s-1}(\ell):=\mathbbm{1}(\ell=1)\lambda_o +  V_{s-1}(\mathcal{F}(\ell))-V_{s-1}(\mathcal{S}(\ell)).
\label{eq:deltadef}
\end{align}
Similarly, let $V^{\pi}_{s}(\ell)$ be the cost incurred by the system starting in state $\ell$ and operating for $s$ time-slots under the application of policy $\pi$. In the following, 
denote by $(u,\pi)$ the policy that chooses the action $u$ in the first time slot irrespective of the initial system state $\ell$, and thereafter implements the policy $\pi$.
\begin{lemma}\label{l2}
For any two actions $u_1,u_2$ and
policy $\pi$,
\begin{align*}
& V^{(u_2,\pi)}_{s}(\mathcal{F}(\ell))-V^{(u_1,\pi)}_{s}(\mathcal{S}(\ell)) \\
& = p(u_1) \left\{ V^{\pi}_{s-1}(\mathcal{S}(\mathcal{F}(\ell)))-V^{\pi}_{s-1}(\mathcal{S}(\mathcal{S}(\ell)))\right\} \\
&\quad \quad +(1-p(u_2)) \left\{  \mathbbm{1}(\mathcal{F}(\ell)=1)\lambda_o - \mathbbm{1}(\mathcal{S}(\ell)=1)\lambda_o \right. \\
&\quad \quad +  V^{\pi}_{s-1}(\mathcal{F}(\mathcal{F}(\ell)))
\left. -V^{\pi}_{s-1}(\mathcal{F}(\mathcal{S}(\ell)))\right\}+\hat{c}(u_2)-\hat{c}(u_1)\\
&\quad \quad + \mathbbm{1}(\mathcal{F}(\ell)=0)- \mathbbm{1}(\mathcal{S}(\ell)=0) \\
& \quad \quad - \left(p(u_2)-p(u_1)\right) \mathbbm{1}(\mathcal{S}(\ell)=1)\lambda_o \\
&= p(u_1) \left\{ V^{\pi}_{s-1}(\mathcal{F}(\mathcal{S}(\ell)))-V^{\pi}_{s-1}(\mathcal{S}(\mathcal{S}(\ell)))\right\} \\
&\quad \quad +(1-p(u_2)) \left\{ \mathbbm{1}(\mathcal{F}(\ell)=1)\lambda_o - \mathbbm{1}(\mathcal{S}(\ell)=1)\lambda_o 
\right. \\
&\quad \quad +  V^{\pi}_{s-1}(\mathcal{F}(\mathcal{F}(\ell))) 
\left. -V^{\pi}_{s-1}(\mathcal{S}(\mathcal{F}(\ell)))\right\}+\hat{c}(u_2)-\hat{c}(u_1)\\
&\quad \quad + \mathbbm{1}(\mathcal{F}(\ell)=0)- \mathbbm{1}(\mathcal{S}(\ell)=0) \\
& \quad \quad - \left(p(u_2)-p(u_1)\right) \mathbbm{1}(\mathcal{S}(\ell)=1)\lambda_o. \\
\end{align*}
\end{lemma}
\begin{IEEEproof}
\begin{align*}
&V^{(u_2,\pi)}_{s}(\mathcal{F}(\ell)) = \hat{c}(u_2)+ \mathbbm{1}(\mathcal{F}(\ell)=0)\\
& + \left(p(u_2)-p(u_1)\right) V^{\pi}_{s-1}(\mathcal{S}(\mathcal{F}(\ell))) \\
&+p(u_1) V^{\pi}_{s-1}(\mathcal{S}(\mathcal{F}(\ell))) \\
&+(1-p(u_2)) \left( \mathbbm{1}(\mathcal{F}(\ell)=1)\lambda_o + V^{\pi}_{s-1}(\mathcal{F}(\mathcal{F}(\ell))) \right), \\
\end{align*}
and
\begin{align*}
&V^{(u_1,\pi)}(\mathcal{S}(\ell)) = \hat{c}(u_1)+ \mathbbm{1}(\mathcal{S}(\ell)=0)\\
&+p(u_1) V^{\pi}_{s-1}(\mathcal{S}(\mathcal{S}(\ell)))  \\
&+(1-p(u_2)) \left( \mathbbm{1}(\mathcal{S}(\ell)=1)\lambda_o + V^{\pi}_{s-1}(\mathcal{F}(\mathcal{S}(\ell))) \right) \\
&+\left(p(u_2)-p(u_1)\right)   \left( \mathbbm{1}(\mathcal{S}(\ell)=1)\lambda_o + V^{\pi}_{s-1}(\mathcal{F}(\mathcal{S}(\ell))) \right).
\end{align*}
Subtracting one from the other, and using $\mathcal{F}(\mathcal{S}(\ell))=\mathcal{S}(\mathcal{F}(\ell))$ in order to cancel the terms that are multiplied by $p(u_2)-p(u_1)$, yields the desired result.
\end{IEEEproof}

\begin{lemma}\label{eq:hjb}
Let us assume that the function $\mathcal{D}_{s-1}(\ell)$ is non-increasing, i.e., $\mathcal{D}_{s-1}(\ell) \geq \mathcal{D}_{s-1}(\ell+ 1)$ for $\ell\in \{1,2,\ldots,B-1\}$.\par 
Let $OPT^{s}(\ell)$ denote the set of actions that are optimal for state $\ell$ for the system starting in state $\ell$ with $s$ time-slots to-go. Then either $OPT^{s}(\ell) = OPT^{s}(\ell+1)$, or we have that if $u_1\in OPT^{s}(\ell+1),u_2\in OPT^{s}(\ell)$ then $p(u_2)>p(u_1)$. In the latter case, the sets $ OPT^{s}(\ell+1), OPT^{s}(\ell)$ have an empty intersection, as a consequence.
\end{lemma}
\begin{IEEEproof}
Consider two actions $u_1,u_2$ such that $u_1 \in OPT^{s}(\ell+1)$, while $u_2\in OPT^{s}(\ell)$. The following inequalities follow from the definition of optimal action~\eqref{eq:1},
\begin{align}\label{eq:monotone1}
\hat{c}(u_2)-p(u_2) \mathcal{D}_{s-1}(\ell+1) &\geq  \hat{c}(u_1)-p(u_1) \mathcal{D}_{s-1}(\ell+1), \notag\\
\hat{c}(u_1)-p(u_1) \mathcal{D}_{s-1}(\ell) &\geq  \hat{c}(u_2)-p(u_2) \mathcal{D}_{s-1}(\ell).
\end{align}
Adding the above two inequalities, we get
\begin{align*}
p(u_1) \left( \mathcal{D}_{s-1}(\ell+1)-\mathcal{D}_{s-1}(\ell)\right) \geq p(u_2) \left( \mathcal{D}_{s-1}(\ell+1)-\mathcal{D}_{s-1}(\ell)\right), 
\end{align*}
or equivalently
\begin{align}\label{eq:ds}
\left(p(u_2)-p(u_1) \right)\left( \mathcal{D}_{s-1}(\ell)-\mathcal{D}_{s-1}(\ell+1)\right) \geq 0. 
\end{align}
Since we assumed that the function $\mathcal{D}_{s-1}(\ell)$ is non-increasing, we have that $ \mathcal{D}_{s-1}(\ell)-\mathcal{D}_{s-1}(\ell+1) \geq 0$. We consider the following two possibilities.\\
\emph{Case A. $ \mathcal{D}_{s-1}(\ell)-\mathcal{D}_{s-1}(\ell+1) = 0$.} \\
We infer the following from the set of inequalities~\eqref{eq:monotone1},
\begin{align*}
\hat{c}(u_2)-p(u_2) \mathcal{D}_{s-1}(\ell+1) &\geq  \hat{c}(u_1)-p(u_1) \mathcal{D}_{s-1}(\ell+1) \\
&=\hat{c}(u_1)-p(u_1) \mathcal{D}_{s-1}(\ell)\\
 &\geq  \hat{c}(u_2)-p(u_2) \mathcal{D}_{s-1}(\ell)\\
 &=\hat{c}(u_2)-p(u_2) \mathcal{D}_{s-1}(\ell+1),
\end{align*}
where the equalities follow from our assumption that $\mathcal{D}_{s-1}(\ell)=\mathcal{D}_{s-1}(\ell+1)$. Thus, the inequalities in the above turn out to be equalities, and we have that 
\begin{align*}
\hat{c}(u_2)-p(u_2) \mathcal{D}_{s-1}(\ell+1) &=  \hat{c}(u_1)-p(u_1) \mathcal{D}_{s-1}(\ell+1) \mbox{ and },\\
\hat{c}(u_1)-p(u_1) \mathcal{D}_{s-1}(\ell) &=  \hat{c}(u_2)-p(u_2) \mathcal{D}_{s-1}(\ell),
\end{align*}
i.e., the actions $u_1,u_2$ are both optimal for the states $\ell,\ell+1$. Since the choice of $u_1\in OPT^{s}(\ell+1),u_2\in OPT^{s}(\ell)$ was arbitrary, we conclude that $OPT^{s}(\ell)=OPT^{s}(\ell+1)$.\\
\\
\emph{Case B. $ \mathcal{D}_{s}(\ell)-\mathcal{D}_{s}(\ell+1) >0$.}\\
It clearly follows from~\eqref{eq:ds} that $p(u_2)>p(u_1)$. \\
This concludes the proof.
\end{IEEEproof}

\begin{lemma}\label{l3}
For each $s=1,2,\ldots$, the function $\mathcal{D}_{s}(\ell)$ defined in~\eqref{eq:deltadef} as
\begin{align*}
\mathcal{D}_{s}(\ell):=\mathbbm{1}(\ell=1)\lambda_o +  V_{s}(\mathcal{F}(\ell))-V_{s}(\mathcal{S}(\ell))
\end{align*}
is non-increasing for $\ell\in \{1,2,\ldots,B\}$. Hence, the properties of the sets $OPT^{s}(\ell)$, which were derived in Lemma~\ref{eq:hjb} under this assumption, are true. 
\end{lemma}
\begin{IEEEproof}
Within this proof, let $\pi_s^{\star}$ be the optimal policy when the time-horizon is $s$ time-slots as in~\eqref{disc}, and let $(u,\pi_{s-1}^{\star})$ be the policy for $s$ time-slots which takes the action $u$ at the first time-slot, and then follows the policy $\pi_{s-1}^{\star}$. We will use induction on $s$, the number of time-slots.

Let us assume that the statement is true for the functions $\mathcal{D}_{z}(\ell)$, for all $z< s$. In particular this implies the function, 
\begin{align}\label{assum1}
\mathcal{D}_{s-1}(\ell) = \mathbbm{1}(\ell=1)\lambda_o +  V_{s-1}(\mathcal{F}(\ell))-V_{s-1}(\mathcal{S}(\ell)), 
\end{align}
is non-increasing for $\ell\in \{1,2,\ldots,B\}$. 

First we will prove the non-increasing property for $\ell\in \{2,3,\ldots,B-T+1\}$.
The above assumption~\eqref{assum1} and Lemma~\ref{eq:hjb} together imply that $\pi_s^{\star}$ is of threshold-type.

 Fix an $\ell \in \{1,2,\ldots,B\}$ and denote by $u_1,u_2,u_3,u_4$, the actions chosen by $\pi^{\star}_s$ at time $s$ for the states $\mathcal{S}(\ell),\mathcal{F}(\ell),\mathcal{S}(\ell+1),\mathcal{F}(\ell+1)$ respectively. Note that the threshold nature of $\pi_s^{\star}$ implies that, 
\begin{align*}
& p(u_1)\leq p(u_2), p(u_3)\leq p(u_4),\mbox{ and }\\
& p(u_3)\leq p(u_1), p(u_4)\leq p(u_2).
\end{align*}
This is true because as the value of state decreases in the interval $\{1,2,\ldots,B\}$, a monotone policy switches to an action that has a higher transmission success probability. 

For $\ell\in \{2,3,\ldots,B-T+1\}$, we have that $\mathcal{D}_{s}(\ell+1)= V_{s}(\mathcal{F}(\ell+1))-V_{s}(\mathcal{S}(\ell+1))$ and $\mathcal{D}_{s}(\ell)= V_{s}(\mathcal{F}(\ell))-V_{s}(\mathcal{S}(\ell))$. Thus,
\begin{align*}
& \mathcal{D}_s(\ell+1) =V_{s}(\mathcal{F}(\ell+1))-V_{s}(\mathcal{S}(\ell+1))\\
&\leq V_{s}^{(u_2,\pi_{s-1}^{\star})}(\mathcal{F}(\ell+1))-V_{s}(\mathcal{S}(\ell+1))\\
& =\hat{c}(u_2) -\hat{c}(u_3) \\
&+ p(u_3)\times \left[V_{s-1}(\mathcal{F}(\mathcal{S}(\ell+1)))-V_{s-1}(\mathcal{S}(\mathcal{S}(\ell+1)))\right]\\
&+\left(1-p(u_2)\right)\times \\
& \left\{\mathbbm{1}(\mathcal{F}(\ell+1)=1)+ V_{s-1}(\mathcal{F}(\mathcal{F}(\ell+1))) \right. \\
&\qquad\qquad\left. - V_{s-1}(\mathcal{S}(\mathcal{F}(\ell+1))) \right\}\\
&\leq \hat{c}(u_2) -\hat{c}(u_3) \\
&+ p(u_3)\times \left[V_{s-1}(\mathcal{S}(\mathcal{F}(\ell)))-V_{s-1}(\mathcal{S}(\mathcal{S}(\ell)))\right]\\
&+\left(1-p(u_2)\right)\times \\
&\left[\mathbbm{1}(\mathcal{F}(\ell)=1)+  V_{s-1}(\mathcal{F}(\mathcal{F}(\ell))) - V_{s-1}(\mathcal{S}(\mathcal{F}(\ell)))\right]\\
&\leq V_{s}(\mathcal{F}(\ell))-V_{s}(\mathcal{S}(\ell))\\
&=\mathcal{D}_s(\ell)
\end{align*}
where the first inequality follows since a sub-optimal action in the state $\mathcal{F}(\ell+1)$ increases the cost-to-go for $s$ time-slots, the second inequality is a consequence of the assumption that the functions $V_{s-1}(\mathcal{F}(\ell))-V_{s-1}(\mathcal{S}(\ell))$ are decreasing in $\ell$, while the last inequality follows from the fact that a sub-optimal action in the state $\mathcal{S}(\ell)$ will increase the cost-to-go for $s$ time-slots. Thus we have proved the monotone decreasing property of $\mathcal{D}_{s+1}(\cdot)$ for $\ell\in \{2,3,\ldots,B-T+1\}$.

Since for the state $\ell=1$, $\mathcal{D}_s(\ell)$ consists of an extra term $\mathbbm{1}(\ell=1)\lambda_o$, it remains to show that $\mathcal{D}_{s}(1)>\mathcal{D}_{s}(2)$.
Once again, let $u_1,u_2,u_3,u_4$ be the optimal actions at stage $s$ for the states $T,0,T+1,1$ respectively. Using the same argument as above (i.e., assuming that the actions taken at time $s$ in the states $T,T+1$ are the same, and the actions taken in the states $0,1$ are the same), it follows that
$\mathcal{D}_{s}(1)-\mathcal{D}_{s}(2)\geq
\left(1+\lambda_o \right)- \left(V_{s}(T)-V_{s}(T+1)\right)$.
However, then $V_{s}(T)-V_{s}(T+1) \leq 1$ 
(since,
for $s$ stages, one may apply the same actions for the system starting in state $T$, as that for a system starting in state $T+1$, and note that the two systems couple at a time-slot $\tau$, when the latter system hits the state $0$; the hitting time is of course random). So,
$\mathcal{D}_{s}(1)-\mathcal{D}_{s}(2)\geq 0$,
and thus we conclude that the function $\mathcal{D}_{s}(\ell)$ is non-increasing for $\ell\in \{1,2,\ldots,B\}$. In order to complete the proof, we notice that for $s=1$, we have,
$\mathcal{D}_{1}(\ell) = \mathbbm{1}(\ell=1)\lambda_o$,
and thus the assertion of Lemma is true for $s=1$.
\end{IEEEproof}
\begin{theorem}\label{t1}
Any optimal policy for the single-client MDP~\eqref{eq:mdp_single} is necessarily monotonic. Furthermore, there is an optimal policy that is of threshold type.
\end{theorem}
\begin{IEEEproof}
We note that the results in Lemma~\ref{eq:hjb} and Lemma~\ref{l3} were derived for finite time horizon $s$. Firstly we note that since the computation of the sets $OPT^{s}(\cdot)$ correspond to the Policy Iteration algorithm~\cite{puterman} in order to solve the MDP $\min_{\pi_n}\bar{C}^{\pi_n}_n+ \lambda_{e}\bar{E}^{\pi_n}_n$, the sets $OPT^{s}(\ell)$ converge as $s\to\infty$~\cite{puterman}. Denote the limiting sets by $OPT^{\infty}(\ell)$.
Since the state and action spaces are finite, the sets $OPT^{s}(\cdot)$ can assume only finitely many values, and hence the sets $OPT^{s}(\ell)$ converge after finite number of steps to the sets $OPT^{\infty}(\ell)$, which also satisfy the properties derived in Lemma~\ref{eq:hjb} and Lemma~\ref{l3}. Thus, if $u_1\in OPT^{\infty}(\ell+1),u_2\in OPT^{\infty}(\ell)$, then $p(u_2)\geq p(u_1)$. Since for an optimal policy, the action taken in each state $\ell$ is necessarily drawn from the set $OPT^{\infty}(\ell)$, it then follows that an optimal policy is necessarily monotonic.

Next, we construct an optimal policy $\pi$ that is of threshold-type. We will group the states $\ell$ on the basis of the set of actions that are optimal when buffer level is $\ell$, i.e. $OPT^{\infty}(\ell)$. Thus, all states $\hat{\ell}$ for which the set of optimal actions $OPT^{\infty}(\hat{\ell})$ is equal to the set $OPT^{\infty}(\ell)$ belong to the same group as that of $\ell$. Now pick an action $u\in OPT^{\infty}(\ell)$, and let $\pi$ apply the action $u$ for all buffer levels that belong to the group corresponding to the set of states $\tilde{\ell}$ that have the set of optimal actions equal to $OPT^{\infty}(\ell)$. It then follows from Lemma~\ref{l3}, and the construction of $\pi$ that if the actions $u_1,u_2$ taken by $\pi$ in the states $\ell + 1,\ell$ are not the same, then $p(u_2)>p(u_1)$. Thus, $\pi$ is of threshold type, and is optimal for the MDP~\eqref{eq:mdp_single}. This completes the proof.
\end{IEEEproof}
\section{Streaming with $M<N$ orthogonal channels}\label{sec:qpa}
In the system considered so far, it was implicitly assumed that there are 
$N$ orthogonal channels available to the AP in case it needs to transmit packets for all clients concurrently. We now consider the problem when the number of orthogonal channels $M <N$.\footnote{One can similarly consider a constraint on peak transmission power. } Thus the AP has to choose actions $u_n(t)=(e_n(t),r_n(t)),t=1,2,\ldots$ for each client $n$ under the constraint that a maximum of $M$($<N$) clients can be served in any time slot $t$, i..e, $\sum_n \mathbbm{1}(e_n(t)>0)\leq M$. Since we do not impose any constraint on the average power utilization, if client $n$ is provided channel access in time-slot $t$, then it is optimal to let $e_n(t) = e_n$, where $e_n$ is the maximum allowable transmission power for client $n$. Thus, without loss of generality, we let $e_n(t)$ assume binary values; $e_n(t)=1$ denotes that a packet was scheduled for client $n$ at time $t$, while $e_n(t)=0$ otherwise. A dynamic optimization with respect to the resolution $r_n(t)$ still needs to be made.

The following CMDP needs to be solved,
\begin{align}
&\min_{\pi} \sum_{n=1}^{N} \bar{C}_n,\label{hardmdp2}\\
&\mbox{s.t.} \sum_{n=1}^{N} e_n(t) \leq M.\label{hardmdp3}
\end{align}
where $\bar{C}_n$ denotes average value of the QoE cost. The above problem involves ``bandit superprocesses"~\cite{Whittle2011Book,mahajan_bandit}. For such Multi Armed Bandit superprocesses, there are no known policies that are guaranteed to have good performance. However, we will now design an appropriate index policy for the setup of bandit superprocesses. 
\begin{definition}[Index Policy]
An index policy maintains $N$ functions $W_n: \{1,2,\ldots,B_n\}\to \mathbb{R}$, where the function $W_n(\cdot)$ maps the state-action pairs of client $n$ to a value in $\mathbb{R}$. At the beginning of each time-slot $t$ the policy assigns the index $W_n(\ell_n(t))$ to the client $n$, and thereafter schedules $M$ clients having the largest values of indices $W_n(\ell_n(t))$.
\end{definition}
\noindent
\emph{Look-ahead Rule based Index Policy:}\\
We briefly describe the look-ahead rule~\cite{bertsekasdp,bertsekas1999rollout} which is a popular technique to obtain efficient dynamic policies. For an average cost MDP that is characterized by a finite state space $\mathcal{X}$, finite action space $\mathcal{A}$, transition probabilities $P(i,a,j), i,j\in\mathcal{X}, a\in\mathcal{A}$, and one-step state-action cost $c(i,a),i\in\mathcal{X},a\in\mathcal{A}$, the following two-step procedure yields a one-step look-ahead policy.  
\begin{enumerate}
\item Begin with a ``base" policy $\pi$ that maps $\mathcal{X}\to \mathcal{A}$. Compute the value function $V_\pi(\cdot)$ and average reward $\bar{J}_{\pi}$ corresponding to policy $\pi$ by solving the following system of $|\mathcal{X}|$ linear equations
$\forall i\in \mathcal{X}$,
\begin{align}\label{lp}
\bar{J}_{\pi} + V_\pi(i) =c(i,\pi(i)) + \sum_{j\in \mathcal{X}}P(i,\pi(i),j)V_{\pi}(j).
\end{align}
\item Now apply one step of the policy improvement algorithm~\cite{Kumar:1986:SSE:40665,puterman} on the policy $\pi$. This generates a new policy, denoted $\tilde{\pi}$, which is obtained by solving the following system of non-linear equations for all $i \in\mathcal{X}$,
\begin{align}\label{pi}
\tilde{\pi}(i) &= \arg\min_{a} [ c(i,a) + \sum_{j\in\mathcal{X}}P(i,a,j) V_{\pi}(j) ].
\end{align}
\end{enumerate}
The policy improvement operator is known to be equivalent to Newton's method applied on the space of policies~\cite{whittleoc}, and its repeated application yields the optimal policy. However, it has been observed in practice that even a single step of policy improvement produces quite efficient policies~\cite{bertsekas1999rollout}. This is possibly due to the fact that Newton's method utilizes the curvature of the fixed point equation at the current estimate in order to converge faster.

We now show how we can derive index policies using the look-ahead principle. We begin by describing the base policy that will be utilized in order to obtain an index policy.

\noindent
\textit{Base policy $\pi$ of interest } \\
We take the base policy $\pi$ to be the policy that at each time $t$ chooses to schedule each client $n$ with a probability $\frac{M}{N}$. If client $n$ gets chosen for scheduling, then the resolution is chosen uniformly at random from the set $\{1,2,\ldots,R_n\}$. Thus, at each time $t$, client $n$ is scheduled a packet of resolution $r$ with a probability $\frac{M}{N}\frac{1}{R_n}$.\footnote{ Note that since there is no constraint on the energy utilization, we allow the base policy to utilize $e_n$ units of energy for packet transmission.} We note that the base policy as described above satisfies the constraint $\sum_n e_n(t)=M$ only on an average, i.e., $\mathbb{E}\left(\sum_n e_n(t)\right) = M$, and hence the base policy is not feasible for the original problem~\eqref{hardmdp2}-\eqref{hardmdp3} of scheduling a maximum of $M$ clients during each time-slot $t$. The infeasibility occurs because the control processes $u_n(t)$ are not coupled via the hard constraint $\sum_n u_n(t)\leq M, t=1,2,\ldots$, and are independent of each other. Though at first look this may seem to be a problem, it
actually offers a huge advantage because this is the precise reason that the look ahead policy generated from it is an index policy.
 While we are generating the look-ahead policy from the base policy, we will necessarily require it to satisfy the constraint $\sum_n e_n(t)=M$.

The following result follows easily from the structure of the base policy $\pi$.
\begin{lemma}
Under the application of the base policy $\pi$ described above, for each client $n$, the control process $u_n(t)$ is i.i.d. across time. Thus, the value function $V_\pi$ 
decomposes into the sum of the value functions of each client $V_{\pi,n},n=1,2,\ldots$, i.e.,
\begin{align}\label{separable}
V_\pi(\ell_1,\ell_2,\ldots,\ell_N) = \sum_{i=1}^{N} V_{\pi,n}(\ell_i).
\end{align}
The value function $V_{\pi,n}(\cdot)$ corresponding to client $n$ can be obtained by solving $B_n$ linear equations~\eqref{lp}, and hence the computational complexity of obtaining functions $\{V_{\pi,n}\}_{n=1}^{N}$ is linear in $N$.
\end{lemma}
\begin{IEEEproof}
The separability property~\eqref{separable} of the value function follows from the i.i.d. nature of the control process for each client, and the fact that the cumulative cost~\eqref{hardmdp2} incurred by the system is 
the sum of the costs incurred by each client. 
\end{IEEEproof}
\begin{theorem}[Look Ahead Index Policy]\label{th:lai}
The look-ahead policy $\tilde{\pi}$ obtained from the base policy $\pi$ described above is an index policy that attaches the following indices to client $n$,
\begin{align}\label{adhoc1}
&W_n(\ell_n(t))\notag\\
&= \min_{u} [ c_n(\ell_n(t),u) + \sum_{j\in \{1,2,\ldots,B_n\}} P(\ell_n(t),u,j)V_{\pi,n}(j) \notag\\
& -c_n(\ell_n(t),0) - \sum_{j\in \{1,2,\ldots,B_n\}} P(\ell_n(t),0,j)V_{\pi,n}(j) ],
 \end{align}
where the action $0$ corresponds to not assigning power to the client. It
then arranges clients in decreasing values of their indices $W_n(\ell_n(t))$, and schedules $M$ clients having the largest $M$ indices. If client $n$ is chosen for scheduling, then the action implemented for it is the action $u$ that attains the maximum in~\eqref{adhoc1}. 
\end{theorem} 
\begin{IEEEproof}
In the optimization problem stated below, the action $u_n$ for client $n$ can assume values from the following set $\{(r,e): r \in \{1,2,\ldots,R_n\}, e\in \{0,1\}\}$. Since the actions with $e=0$ correspond to not transmitting a packet, we will usually denote them simply by $u=0$. Fix the base policy $\pi$ to be the policy that at each time-slot $t$, picks the client $n$ with probability $M\slash N$, and thereafter transmits resolution $r$ packet w.p. $1\slash R_n$.

It then follows from relation~\eqref{pi} that the following optimization problem needs to be solved in order to obtain $\tilde{\pi}(x)$, i.e.,  the action that look-ahead policy takes while system state is equal to $x=(\ell_1,\ell_2,\ldots,\ell_N)$, 
\begin{align*}
\min_{u= \{u_n\}_{n=1}^{N}:\sum_{n=1}^{N} e_n \leq M} \sum_{n=1}^{N} c_n(\ell_n,u_n) +  P(\ell_n,u_n,j) V_{\pi,n}(j),
\end{align*}
which can equivalently be posed as,
\begin{align*}
\min_{ \vspace{.5cm} u:\sum_{n=1}^{N} e_n \leq M  } \sum_{n=1}^{N} c_n(\ell_n,u_n) + \sum_{j\in \mathcal{X}_n} P(\ell_n,u_n,j) V_{\pi,n}(j) \\
- c_n(\ell_n,0) - \sum_{j\in \mathcal{X}_n} P(i,0,j)V_{\pi,n}(j) .
\end{align*}
The problem above is equivalent to $\max_{u:\sum_{n=1}^N e_n \leq M} \sum_{n=1}^{N} W_n(\ell_n)$, which is solved by picking $M$ clients with the largest indices $W_n(\ell_n)$ given by~\eqref{adhoc1}, and setting the corresponding action $u_n = (1,r)$, where $r$ is given by the value that attains the maximum in~\eqref{adhoc1}.
\end{IEEEproof}
\section{Simulations}\label{sec:simu}
We now present the results of simulation studies to assess the performance of the designed policies.
\subsection{Performance of Optimal Policy for Scheduling under the Average Power Constraint}\label{subsec:averagepower}
We perform simulations to assess the performance of the streaming policy 
of Theorem \ref{decentralization} that is optimal under average power constraint on the AP~\eqref{costavg}-\eqref{pmdp}. 

\begin{figure}[h]
\includegraphics[width=8cm]{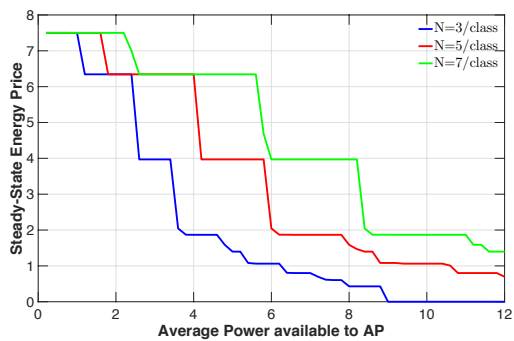}
\caption{Plot of steady-state energy price as the average power at the AP is varied, for the network shown in Table~\ref{table1}, for different numbers of clients per class. }
\label{s1}
\end{figure}

There are 3 classes of clients, with the parameters for the three classes as shown in Table~\ref{table1}. The buffer size is $B=20$, and a single packet is played for $T=5$ time-slots for all the clients. For a fixed set of clients we vary the average power available to the AP, and plot the steady-state link prices that are obtained while solving the linear program corresponding to the CMDP~\eqref{costavg}-\eqref{pmdp}. Fig.~\ref{s1} shows the variations in the steady-state optimal energy price $\lambda^{\star}$ as the number of clients per class is varied. We observe that for a fixed number of clients, the energy price decreases with the available power, while for a fixed value of available power, it increases with the number of clients.

\begin{figure}[h]
\includegraphics[width=8cm]{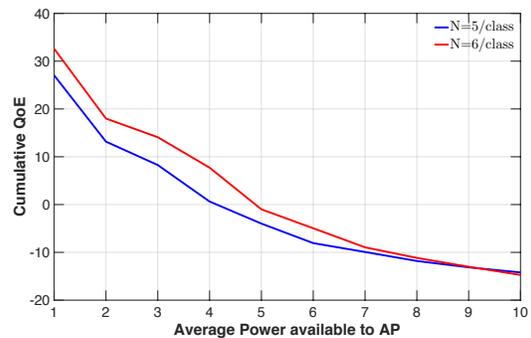}
\caption{Plot of cumulative QoE as the average power at AP is varied for the network shown in Table~\ref{table1} for the case of $5$ and $6$ clients per class.}
\label{s5}
\end{figure}

We plot the cumulative QoE of the system as a function of the available average power at the AP in Fig.~\ref{s5}.
\begin{table}[h]
\begin{tabular}{|l|l|l|l|}
\hline 
  Class $1$ & $e_1=.50$ & $e_1=.75$  \\\hline
$v_1=1$ & $.5$& $.6$\\\hline
$v_1=1.5$ &  $.45$& $.55$\\\hline
\end{tabular}
\hfill
\begin{tabular}{|l|l|l|l|}
\hline 
 Class $2$  & $e_2=.75$ & $e_2=1$  \\\hline
$v_1=1.5$ & $.75$& $.85$\\\hline
$v_1=2.0$ &  $.70$& $.80$\\\hline
\end{tabular}
\hfill
\begin{tabular}{|l|l|l|l|}
\hline 
 Class $3$  & $e_3=.85$ & $e_3=1.2$  \\\hline
$v_3=.75$ & $.65$& $.75$\\\hline
$v_3=1.5$ &  $.60$& $.70$\\\hline
\end{tabular}
\end{table}
\begin{table}[h]
\caption{
The entry in the cell at intersection of $e_i,v_i$ is the probability of successful packet transmission for a client of class $i$ with the given values of transmission power and resolution. 
}
\label{table1}
\end{table}
\subsection{Look-Ahead Index Policy}
We now assess the performance of the look ahead based index policy derived in Theorem~\ref{th:lai} for scheduling clients when $M$ ($<N$) orthogonal channels are available for concurrent packet transmissions, and the AP is allowed to use resolution-power adaptation. As in previous sections, a client can belong to one of the three classes. The quantities $VQ(i,j),P(i,j)$ in~\eqref{eq:laiparameters} are the resolution and transmission probability associated with the $j$-th resolution video of the $i$-th class. Similarly, the $i$-th entry of $B,T$ vectors denote the buffer size and play time for $i$-th class.
\begin{align}\label{eq:laiparameters}
VQ &= \begin{bmatrix}-2 &-3 &-4 &-5 \\ -1 &-3 &-5& -5.5\\ -1 &-2.5 &-3 &-4\end{bmatrix}, B = \begin{bmatrix} 10&8 &6\end{bmatrix},\notag \\
P &= \begin{bmatrix} .8 &.7& .6 &.5 \\.7 &.6 &.5 &.4\\.75& .65 &.45 &.3 \end{bmatrix}, \qquad~ T = \begin{bmatrix}3&4 &5\end{bmatrix}.
\end{align}
\begin{figure}[h]
\includegraphics[width=8cm]{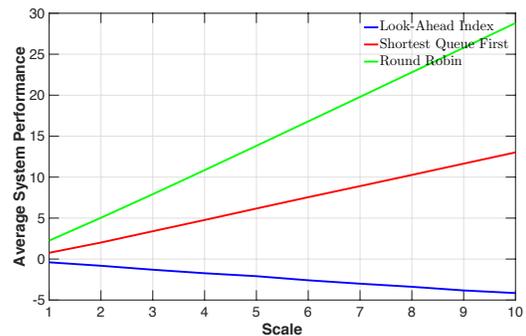}
\caption{Performance of the policies for the network with parameters $VQ,P,B,T$ as in~\eqref{eq:laiparameters}, as the network size is scaled.}
\label{s3}
\end{figure}
\begin{figure}[h]
\includegraphics[width=8cm]{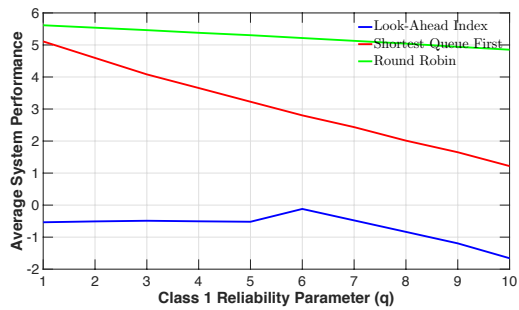}
\caption{Performance of the policies as the channel reliability of class $1$ clients is varied according to $P(1,1)=.4+.05r, P(1,2)=.3+.05r,P(1,3) = .2+.05r,P(1,4)= .1+.05r$, where $r$ is the reliability parameter. }
\label{s4}
\end{figure}
Figure~\ref{s3} shows the performance of the policies as the system size is scaled. We now vary the channel reliability of class $1$ clients, while keeping the other parameters fixed according to the matrices $VQ,P,B,T$ defined above. As shown in Fig.~\ref{s4}, we observe that since resolution-power adaptation requires the optimal decision process to be quite complex, the index policy of Theorem~\ref{th:lai} performs significantly better than the SQF and RR policies. 
\section{Concluding Remarks}\label{sec:con}
We have addressed the problem of designing decentralized scheduling algorithms that maximize the cumulative Quality of Experience of multiple clients streaming video over unreliable channels. When the AP is constrained by its average power, we have shown
that a decentralized policy where the AP charges a price $\lambda^{\star}$ per unit amount of transmission energy, and clients optimize their individual costs is optimal. The price $\lambda^{\star}_e$ solves the Dual Problem~\eqref{dualprob}, and is the price at which the net rate of energy consumption is equal to the available power. The decentralized optimal policy can be obtained by solving a linear program in which the number of variables scales linearly with the number of clients. An iterative algorithm that computes the optimal policy in a distributed manner, is also proposed. It has also been shown that the optimal policy for the single-client MDP is of threshold type. 

When the AP is limited by the number of $M$ ($<N$) orthogonal channels, we have derived index policies. For the set-up without resolution-power adaptation, the scheduling problem is indexable. Indexability of the scheduling problem relies on the result that a threshold policy is optimal for each single client MDP. When the AP is allowed to choose from several power-resolution levels, the problem of scheduling clients can be posed as a multi-armed bandit superprocess. 
We have utilized the one-step look ahead rule/policy improvement on a naive base scheduling policy and showed that the resulting policy is an index policy. The resultant index policies 
are seen to perform well in simulations, with the look-ahead index policy performing much better than the Shortest Queue and
Round Robin policies.

\bibliographystyle{IEEEtran}
\bibliography{IEEEabrv,journal.bib}
\end{document}